\documentclass[12pt]{article}
\usepackage{epsfig}
\usepackage{amssymb,amsmath}
\usepackage{epstopdf}
\hoffset= - 0.8in         

\catcode`\@=11
\def\marginnote#1{}

\newcount\hour
\newcount\minute
\newtoks\amorpm
\hour=\time\divide\hour by60
\minute=\time{\multiply\hour by60 \global\advance\minute by-\hour}
\edef\standardtime{{\ifnum\hour<12 \global\amorpm={am}%
        \else\global\amorpm={pm}\advance\hour by-12 \fi
        \ifnum\hour=0 \hour=12 \fi
        \number\hour:\ifnum\minute<10 0\fi\number\minute\the\amorpm}}
\edef\militarytime{\number\hour:\ifnum\minute<10 0\fi\number\minute}

\def\draftlabel#1{{\@bsphack\if@filesw {\let\thepage\relax
   \xdef\@gtempa{\write\@auxout{\string
      \newlabel{#1}{{\@currentlabel}{\thepage}}}}}\@gtempa
   \if@nobreak \ifvmode\nobreak\fi\fi\fi\@esphack}
        \gdef\@eqnlabel{#1}}
\def\@eqnlabel{}
\def\@vacuum{}
\def\draftmarginnote#1{\marginpar{\raggedright\scriptsize\tt#1}}

\def\draft{\oddsidemargin -.5truein
        \def\@oddfoot{\sl preliminary draft \hfil
        \rm\thepage\hfil\sl\today\quad\militarytime}
        \let\@evenfoot\@oddfoot \overfullrule 3pt
        \let\label=\draftlabel
        \let\marginnote=\draftmarginnote
   \def\@eqnnum{(\theequation)\rlap{\kern\marginparsep\tt\@eqnlabel}%
\global\let\@eqnlabel\@vacuum}  }

\def\d{\partial}

\def\bea{\begin{eqnarray}}
\def\eea{\end{eqnarray}}

\def\beq{\begin{equation}}
\def\eeq{\end{equation}}
\def\ba{\beq\new\begin{array}{c}}
\def\ea{\end{array}\eeq}
\def\be{\ba}
\def\ee{\ea}
\def\stackreb#1#2{\mathrel{\mathop{#2}\limits_{#1}}}
\def\Tr{{\rm Tr}}
\def\Ad{{\rm Ad}}
\def\sign{{\rm sign}}

\makeatletter
\newdimen\normalarrayskip              
\newdimen\minarrayskip                 
\normalarrayskip\baselineskip
\minarrayskip\jot
\newif\ifold             \oldtrue            \def\new{\oldfalse}
\def\arraymode{\ifold\relax\else\displaystyle\fi} 
\def\eqnumphantom{\phantom{(\theequation)}}     
\def\@arrayskip{\ifold\baselineskip\z@\lineskip\z@
     \else
     \baselineskip\minarrayskip\lineskip2\minarrayskip\fi}
\def\@arrayclassz{\ifcase \@lastchclass \@acolampacol \or
\@ampacol \or \or \or \@addamp \or
   \@acolampacol \or \@firstampfalse \@acol \fi
\edef\@preamble{\@preamble
  \ifcase \@chnum
     \hfil$\relax\arraymode\@sharp$\hfil
     \or $\relax\arraymode\@sharp$\hfil
     \or \hfil$\relax\arraymode\@sharp$\fi}}
\def\@array[#1]#2{\setbox\@arstrutbox=\hbox{\vrule
     height\arraystretch \ht\strutbox
     depth\arraystretch \dp\strutbox
     width\z@}\@mkpream{#2}\edef\@preamble{\halign
\noexpand\@halignto
\bgroup \tabskip\z@ \@arstrut \@preamble \tabskip\z@ \cr}%
\let\@startpbox\@@startpbox \let\@endpbox\@@endpbox
  \if #1t\vtop \else \if#1b\vbox \else \vcenter \fi\fi
  \bgroup \let\par\relax
  \let\@sharp##\let\protect\relax
  \@arrayskip\@preamble}
\def\eqnarray{\stepcounter{equation}%
              \let\@currentlabel=\theequation
              \global\@eqnswtrue
              \global\@eqcnt\z@
              \tabskip\@centering
              \let\\=\@eqncr
 \halign to \displaywidth\bgroup
    \eqnumphantom\@eqnsel\hskip\@centering
    $\displaystyle \tabskip\z@ {##}$%
    \global\@eqcnt\@ne \hskip 2\arraycolsep
         $\displaystyle\arraymode{##}$\hfil
    \global\@eqcnt\tw@ \hskip 2\arraycolsep
         $\displaystyle\tabskip\z@{##}$\hfil
         \tabskip\@centering
    &{##}\tabskip\z@\cr}
\begingroup\ifx\undefined\newsymbol \else\def\input#1 {\endgroup}\fi
\newfont{\hr}{msbm10}
\newfont{\ams}{msam10}

\font\numbers=cmss12
\font\upright=cmu10 scaled\magstep1
\def\stroke{\vrule height8pt width0.4pt depth-0.1pt}
\def\topfleck{\vrule height8pt width0.5pt depth-5.9pt}
\def\botfleck{\vrule height2pt width0.5pt depth0.1pt}
\def\Zmath{\vcenter{\hbox{\numbers\rlap{\rlap{Z}\kern 0.8pt\topfleck}\kern
2.2pt
                   \rlap Z\kern 6pt\botfleck\kern 1pt}}}
\def\Qmath{\vcenter{\hbox{\upright\rlap{\rlap{Q}\kern
                   3.8pt\stroke}\phantom{Q}}}}
\def\Nmath{\vcenter{\hbox{\upright\rlap{I}\kern 1.7pt N}}}
\def\Cmath{\vcenter{\hbox{\upright\rlap{\rlap{C}\kern
                   3.8pt\stroke}\phantom{C}}}}
\def\Rmath{\vcenter{\hbox{\upright\rlap{I}\kern 1.7pt R}}}
\def\Z{\ifmmode\Zmath\else$\Zmath$\fi}
\def\Q{\ifmmode\Qmath\else$\Qmath$\fi}
\def\N{\ifmmode\Nmath\else$\Nmath$\fi}
\def\C{\ifmmode\Cmath\else$\Cmath$\fi}
\def\R{\ifmmode\Rmath\else$\Rmath$\fi}

\def\stackreb#1#2{\mathrel{\mathop{#2}\limits_{#1}}}
\def\Tr{{\rm Tr}}

\def\Bf#1{\mbox{\boldmath $#1$}}

\def\bbeta{{\Bf\beta}}

\def\bxi{{\Bf\xi}}
\def\bbeta{{\Bf\eta}}

\def\d{\partial}

\def\rank{{\rm rank}}

\def\half{{\textstyle{1\over2}}}
\def\quatr{{\textstyle{1\over4}}}
\def\2{{1\over 2}}

\def\beq{\begin{equation}}
\def\eeq{\end{equation}}
\def\ba{\beq\new\begin{array}{c}}
\def\ea{\end{array}\eeq}
\def\be{\ba}
\def\ee{\ea}
\def\stackreb#1#2{\mathrel{\mathop{#2}\limits_{#1}}}
\def\theequation{\thesection.\arabic{equation}}

\newcommand{\rf}[1]{(\ref{#1})}

\begin{document}


\renewcommand{\thefootnote}{\fnsymbol{footnote}}
\begin{center}
\baselineskip20pt
{\bf \LARGE Lie Groups, Cluster Variables\\ and Integrable Systems}
\end{center}
\bigskip
\begin{center}
\baselineskip12pt
{\large A.~Marshakov\footnote{Based on the talks given at Versatility of integrability, Columbia University, May 2011; Simons Summer Workshop on Geometry and Physics, Stony Brook, July-August 2011; Classical and Quantum Integrable Systems, Dubna, January 2012; Progress in Quantum Field Theory and String Theory, Osaka, April 2012; Workshop on Combinatorics of Moduli Spaces and Cluster Algebras, Moscow, May-June 2012.}}\\
\bigskip
{\em Lebedev Physics Institute, ITEP and NRU-HSE, Moscow, Russia}\\
\medskip
{\sf e-mail:\ mars@lpi.ru, mars@itep.ru}\\
\end{center}
\bigskip\medskip

\begin{center}
{\large\bf Abstract} \vspace*{.2cm}
\end{center}

\begin{quotation}
\noindent
We discuss the Poisson structures on Lie groups and
propose an explicit construction of the integrable models on their appropriate Poisson submanifolds.
The integrals of motion for the $SL(N)$-series are computed
in cluster variables via the Lax map. This construction, when generalised to the co-extended loop
groups, gives rise not only to several
alternative descriptions of relativistic Toda systems, but allows to formulate in general terms some new class of integrable models.
\end{quotation}

\renewcommand{\thefootnote}{\arabic{footnote}}
\setcounter{section}{0}
\setcounter{footnote}{0}
\setcounter{equation}0
\section{Introduction}

This is a review and announcement of some results, mostly obtained in collaboration with
Vladimir Fock \cite{FM,FM11,FM12} and yet unpublished. The relation between integrable systems and Lie groups has many
different faces, see e.g. \cite{OP}, but following
the old proposal of \cite{FM}, an integrable model can be {\em directly}
constructed on Poisson submanifold in Lie group. Integrability in this case
just follows from existence of the Ad-invariant functions on group
manifold. Going further, it has been found \cite{FM11},
that this idea can be
extended to affine Lie groups and gives
rise to a nontrivial wide class of integrable systems \cite{FM12}, which has been alternatively
discovered in \cite{GK} as arising from the dimer models on bipartite graphs on a torus.
These results are partially intersecting and complementary with presented in \cite{OT,GeSha,GeShaLast}, and other references to closely related works include \cite{Sol,eager,will}.

For the canonical Poisson structure on simple Lie group $G$
one immediately gets
 $\mbox{rank}G$ mutually Poisson-commuting functions.
Restricting them to a symplectic leaf of dimension $2\ \rank G$ we obtain
a completely integrable system, which turns to be equivalent \cite{FM}
to relativistic non-periodic Toda chain \cite{Ruj}. An effective way to construct
the corresponding Poisson submanifold implies using the cluster co-ordinates on Lie groups
\cite{cv,Drinfeld}. The cluster language allows to generalize the old proposal of \cite{FM}
to the case of loop groups and periodic Toda system, requiring the corresponding loop group
to be co-extended \cite{FM11}.
This also opens an opportunity to go far beyond this simple relation between
Lie groups and integrable system we start with, and to develop a new class
of integrable models starting with cluster varieties \cite{FM12}, which is also important
from the point of view of relation with geometry and physics.

In sect.~\ref{ss:isrmatr} we describe the Poisson submanifolds in Lie groups in terms of
the cluster variables. In sect.~\ref{ss:reltoda} we construct explicitly the integrable system -
relativistic Toda chain - both for the cases of simple and co-extended loop groups. The most
nontrivial part of the construction is formulated in sect.~\ref{ss:periodic} and contains
localisation procedure for the co-extended loop group, which allows to fix the ambiguity
in the spectral curve equation. In sect.\ref{ss:nN} the construction of
relativistic Toda chain is generalised to the wider class of integrable models.

\setcounter{equation}0
\section{Poisson brackets and $r$-matrices
\label{ss:isrmatr}}

Let $\mathfrak{g}$ be the Lie algebra of a simple group $G$,  $r\in \mathfrak{g}\otimes \mathfrak{g}$ - the standard solution to the (modified) Yang-Baxter equation, which
defines a Poisson bracket on the group $G$ by
\be\label{rbra}
\{ g \stackreb{,}{\otimes}g \} = -\half\ [r,g\otimes g]
\ee
which is compatible with the group structure
(see Appendices~\ref{ap:promunu} and \ref{ap:rmatr} for notations and some
extra formulas concerning $r$-matrices for the $SL(N)$-series).

The Poisson bracket \rf{rbra} is degenerate on the whole group, but can be restricted on any
symplectic leaf of $G$. A Lie group $G$ can be decomposed into the set of Poisson submanifolds
(see for example \cite{sympleaves,HKKR,Drinfeld}), labeled by $u\in W\times W$, where $W$ is the Weyl group of $G$. The dimension of submanifold $G^u$ is
$
l(u)+\rank G$, where $l(u)$ is the length of the word $u$, which consists of the generators of the Weyl group $W$ (to be identified with the set $\Pi$ of the positive simple roots of $G$), and the generators of the second copy of $W$ (to be identified with the set of negative simple roots ${\bar\Pi}$).
For our purposes it is more
convenient to consider similar decomposition of the factor $G/\mbox{Ad}H$ over the Cartan subgroup $H\subset G$, the dimensions of the corresponding symplectic leaves in the factor are just
the lengths $l(u)$ themselves.

One can construct \cite{Drinfeld} a parameterisation of $G^u/\mbox{Ad}H$,
such that the Poisson bracket \rf{rbra} becomes logarithmically constant:
for any reduced decomposition $u=\alpha_{i_1}\ldots \alpha_{i_{l}}$ consider
the Lax map
\be
\label{paramG/AdH}
z_1,\ldots,z_l \mapsto E_{i_1}H_{i_1}(z_1)\cdots E_{i_l}H_{i_l}(z_l)
\ee
For the group $SL(N)$ in the r.h.s. one can just substitute the matrices ($i=1,\ldots,N-1$)
\be
\label{genslN}
H_i(z) = \left(\begin{array}{cccccc}
z&0&&\cdots&&0\\
0&\ddots&&&&0\\
&&z&&\\
\vdots&&&1&\\
&&&&\ddots&0\\
0&&\cdots&&0&1
\end{array}\right),\ \ \ \
E_i =
\left(\begin{array}{cccccc}
1&0&&\cdots&&0\\
0&\ddots&&&&0\\
&&1&1&\\
\vdots&&&1&\\
&&&&\ddots&0\\
0&&\cdots&&0&1
\end{array}\right)
\ee
where the last line in diagonal $H_i(z)$ with $z\neq 1$ and the only line in $E_i$ containing the off-diagonal unity have number $i$.
For negative $i$ the corresponding matrix is just transposed to the matrix of positive root, i.e.
$E_{\bar i}=E_{-i}=E_{i}^{\rm tr}$. Matrix $H_i(z)\in PGL(N)$ should be
normalized for the $SL(N)$ group $H_i(z)\to Z_i = H_i(z)/(\det H_i(z))^{1/N}$, and therefore will depend always on the fractional powers $z^{k/N}$ of the variable $z$.

The Poisson brackets \rf{rbra} on the parameters $z_I$ is log-constant and half-integral, i.e.
\be
\label{pbclust}
\{z_I,z_J\}=\varepsilon_{IJ}z_Iz_J
\ee
 where $\varepsilon_{IJ}$ is a skew-symmetric matrix taking integral or half-integral values,
and playing a role of the exchange matrix (if the group $G$ is simply laced, e.g. $SL(N)$) for the corresponding cluster variety \cite{cv}. All formulas for symplectic leaves in $G/\mbox{Ad}H$ can be read from a graph with the oriented edges, see Appendix~\ref{ap:poisson}.
For the Poisson submanifolds in $G/\mbox{Ad}H$ one just needs to turn to the graphs on a cylinder instead of a plane. As was already pointed out, among all $G^u$ of particular interest are the cells, corresponding to
the Coxeter elements of $W$ with $l(u)=2\cdot\rank\ G$.

For the loop groups $\widehat{G}$ one gets infinitely many  $\Ad$-invariant functions, but they still posses finite-dimensional Poisson submanifolds, thus providing naturally to construction of wider classes of integrable systems \cite{FM11}. However,
since the Cartan matrices for affine groups \rf{CslNh} are non-invertible, for the cluster construction
one has to use instead the co-extended version of loop group $\widehat{G^\sharp}$, see \cite{Kac}.

The group $\widehat{SL(N)^\sharp}$ can be identified \cite{FM11,FM12} with the group
\be
\label{AshifT}
A_1(\lambda)T_{z_1}\cdot A_2(\lambda)T_{z_2}=A_1(\lambda)A_2(z_1\lambda)T_{z_1z_2}
\ee
of expressions $A(\lambda)T_z$, where
\be
\label{shiftT}
T_z = \exp\left({\log z\frac{\partial}{\partial \log\lambda}}\right) = z^{\lambda\d/\d \lambda}
\ee
and $A(\lambda)$ is a Laurent polynomial with values in $N\times N$ matrices.
The generators of the co-extended group ($i\in\mathbb{Z}_N$) have the form
\be
\label{notloop}
{\bf H}_i(z) =
H_i(z)T_z,\ \ \ \
\mathbf{E}_i =E_i,\ \ \ \ \mathbf{E}_{\bar i} =E_i^{\rm tr}
\ee
for positive $i\neq 0$, i.e. each corresponding root generator just coincides with \rf{genslN},
while each Cartan generator is multiplied by the shift operator.
For $i=0$ we have additionally
\be
\label{loopextra}
\mathbf{H}_0(z)=T_z,\ \  \mbox{ and }\ \
\mathbf{E}_0 = \left(\begin{array}{ccc}
1&\cdots&0\\
\vdots&\ddots&\vdots\\
\lambda&\cdots&1
\end{array}\right)
\ee
It is also useful to introduce the element $\Lambda \in \widehat{SL(N)}$ (as usual, up to
normalisation)
\be
\label{lpshft}
\Lambda =
\left(\begin{array}{cccc}
0&1&\cdots&\lambda^{-1}\\
\vdots&\ddots&\ddots&\vdots\\
0&\cdots&0&1\\
0 &\cdots&0&0
 \end{array}\right)
\ee
with the property
\be
\label{ddshft}
\Lambda \mathbf{E}_i \Lambda^{-1}= \mathbf{E}_{i+1},\ \ \
\Lambda \mathbf{H}_i(z) \Lambda^{-1}= \mathbf{H}_{i+1}(z),\ \ \ i\in{\mathbb{Z}_N}
\ee
i.e. this operator acts as a unit shift along the Dynkin diagram. It can be also interpreted as a coextension of the affine Weyl group $\widehat{W}^\sharp$, for detailed discussion of this issue see \cite{FM12}.

One can consider now arbitrarily long words, corresponding to the Poisson submanifolds in $\widehat{SL(N)^\sharp}$  of arbitrary large dimensions. However, the space of $\Ad$-invariant functions is infinite only for $\widehat{SL(N)}$, but not for the co-extended group. Therefore, to get sufficiently many independent $\Ad$-invariant functions, one has to consider similarly to \rf{paramG/AdH} the Lax maps
\be
\label{paramGloop}
z_1,\ldots,z_l \mapsto \mathbf{E}_{j_1}\mathbf{H}_{j_1}(z_1)\cdots \mathbf{E}_{j_l}\mathbf{H}_{j_l}(z_l)
\ee
where the parameters $z_I$ must satisfy
\be
\label{trishift}
\prod_j z_j = 1,\ \ \ \ \ \ \prod_j T_{z_j} = T_{\prod_j z_j} = {\rm Id}
\ee
i.e. the total co-extension is trivial.

It is easy to understand, that exchange graphs for symplectic leaves in $SL(N)$ and Poisson submanifolds in loop groups $\widehat{SL(N)}$
\begin{figure}[hc]
\epsfysize=5cm
\centerline{\epsfbox{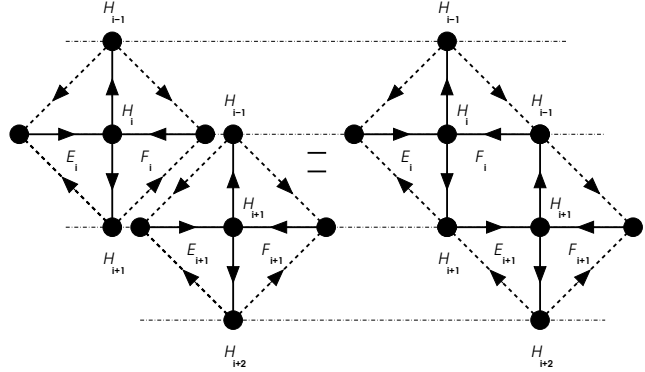}}
\caption{\sl Graphs, depicting the subgroups of $SL(N)$ generated by the positive simple root $E_i$ and negative root $E_{\bar i}$, $1<i<N-1$, or the subgroups of $\widehat{SL(N)}$ for $i\in \mathbb{Z}_N$.
Gluing two such graphs one gets an element of the two-dimensional
square lattice.}
\label{fi:lattice}
\end{figure}
can be constructed by gluing the rhombi (see fig.~\ref{fi:lattice} and Appendix~\ref{ap:poisson})
on a cylinder - for the simple group and on a torus - for the loop group case.
 The graphs for $SL(N)$ would have $N-1$ vertical levels (the corresponding higher and lower rhombi are cut into triangles) and glued in the horisontal direction, while the
graphs for the Poisson submanifolds of the same dimension in $\widehat{SL(N)}$ would have one extra level
(for the same $N$) and will be periodic in vertical direction as well. This construction will be illustrated by explicit examples below.

\setcounter{equation}0
\section{Integrable system
\label{ss:reltoda}}

It is easy to see directly from \rf{rbra}, that any two $\Ad$-invariant functions on $G$ do Poisson-commute with each other. For example, in the simply-laced case, when the $r$-matrix is
\be
\label{ref}
r =  \sum_{\alpha\in\Delta _+} e_\alpha \wedge e_{\bar\alpha} =
\sum _{\alpha\in\Delta _+}\left( e_{\alpha}\otimes e_{\bar\alpha} -
e_{\bar\alpha}\otimes e_{\alpha}\right)
\ee
(the sum is taken over the set of all positive roots), the Poisson bracket is given by
\be
\label{gpb}
\{\mathcal{H}_1,\mathcal{H}_2\}=-\half\sum_{\alpha\in\Delta _+}\left(L_{e_\alpha}\mathcal{H}_1L_{e_{\bar\alpha}}\mathcal{H}_2-
R_{e_\alpha}\mathcal{H}_1R_{e_{\bar\alpha}}\mathcal{H}_2\right)
\ee
where for any $v\in \mathfrak{g}$ we denote by $L_v$ ($R_v$) the corresponding left (right) vector field.
Any $\Ad$-invariant function $\mathcal{H}$ satisfies $L_v \mathcal{H} = -R_v \mathcal{H}$ and thus the bracket \rf{gpb} of two such functions vanishes.
The bracket \rf{gpb} obviously vanishes even if the functions are defined not on the whole $G$, but on any Poisson $\Ad$-invariant subvariety of $G$.

On a simple group there exists $\rank\ G$ independent $\Ad$-invariant functions: a possible basis of these functions is the set $\{\mathcal{H}_i\}$, where $i \in \Pi$  (the set of simple roots)
\be
\label{trfund}
\mathcal{H}_i = \Tr\ \pi_{\mu^i}(g)
\ee
and $\pi_{\mu^i}$ be the $i$-th fundamental representation of $G$ with the highest weight $(\mu^i,\alpha_j)=\delta^i_j$ dual to $\alpha_i$, $i\in\Pi$. These function then define an integrable system on symplectic leaf of dimension $2\cdot \rank\ G$.
It has been shown \cite{FM}, that they
form the set of integrals of motion for the open relativistic Toda chain \cite{Ruj}, with
the canonical Hamiltonian
\be
\label{HslN}
{\cal H} = \Tr\left(g+g^{-1}\right) = \sum_{i=1}^N\left(\exp(p_i)+\exp(-p_i)\right)\sqrt{1+\exp(q_i-q_{i+1})}\sqrt{1+\exp(q_{i-1}-q_i)}
\ee
while
more well-known (non-relativistic) Toda system is recovered in the limit from the Lie group to Lie algebra, see Appendix~\ref{ap:alg}.

\subsection{Simple group: examples}

{\bf SL(2)}.
On symplectic leaf of 2-particle relativistic Toda, corresponding to the word $u=1{\bar 1}$, one can choose the following parameterisation
\be
\label{gxy}
g(x,y) \simeq Y^{1/2}EX{\bar E}Y^{1/2} \simeq YEX{\bar E}
\simeq EX{\bar E}Y =
{1\over \sqrt{xy}}\left(\begin{array}{cc}
xy+y&1\\y&1\end{array}\right)
\ee
where
\be
\label{yxdefa}
Y = H(y)/\det H(y)^{1/2} = \left(
\begin{array}{cc}
  y^{1/2} &  0  \\
  0 & y^{-1/2}
\end{array}\right),
\\
X = H(x)/\det H(x)^{1/2} = \left(
\begin{array}{cc}
  x^{1/2} & 0 \\
 0  & x^{-1/2}
\end{array}\right)
\\
E = \left(
\begin{array}{cc}
  1 & 1 \\
  0 & 1
\end{array}\right),\ \ \
{\bar E} = \left(
\begin{array}{cc}
  1 & 0\\
 1  & 1
\end{array}\right)
\ee
Expansion \rf{gxy} (refinement of the Gauss decomposition) corresponds to the graph
$y\rightarrow x\leftarrow y$ with three vertices and two edges. Identifying the ends, it turns into $y\Rightarrow x$, inducing the Poisson structure
\be
\label{brasl2}
 \{ y,x\} = 2yx
\ee
The cluster variables are related to the Darboux
co-ordinates by
\be
x = e^{-q},\ \ \
y = {e^{2p}\over 1+x}
\ee
while the (only in this example) Hamiltonian is just
\be
\label{hamsl2}
\mathcal{H} = \Tr\ g(x,y) = \sqrt{xy} + \sqrt{y\over x} + {1\over\sqrt{xy}} = \left(e^p+e^{-p}\right)\sqrt{1+e^q}
\ee
the canonical Hamiltonian \rf{HslN} for the 2-particle open system.

\bigskip\noindent
{\bf SL(3)}. For the group $SL(3)$ the formula \rf{paramG/AdH}
gives
\be
\label{gauss3}
g({\bf x},{\bf y})  =
\underbrace{E_1X_1E_{\bar 1}Y_1}_{g_1}\cdot
\underbrace{E_2X_2E_{\bar 2}Y_2}_{g_2}
\ee
where
\be
\label{SL3}
E_1 = E_{\bar{1}}^{\rm tr}=\left(\begin{array}{ccc}
1&1&0\\0&1&0\\0&0&1
\end{array}\right),\ \ \ \
E_2 = E_{\bar{1}}^{\rm tr}=
\left(\begin{array}{ccc}
1&1&0\\0&1&0\\0&0&1
\end{array}\right)
\ee
and the normalised Cartan elements are
\be
\label{yx3def}
Y_i = H_i(y_i)/\det H_i(y_i)^{1/3},\ \ \
X_i = H_i(x_i)/\det H_i(x_i)^{1/3},\ \ \ i=1,2
\\
 H_1(z)=
\left(\begin{array}{ccc}
z&0&0\\
0&1&0\\
0&0&1
\end{array}\right),\ \ \ \
H_2(z)=
\left(\begin{array}{ccc}
z&0&0\\
0&z&0\\
0&0&1
\end{array}\right)
\ee
The variables ${\bf x},{\bf y}$ correspond to the vertices of the graph ${ }^{x_1}_{y_2}\downarrow^{\Leftarrow}_{\Rightarrow}\uparrow{ }^{y_1}_{x_2}$,
\begin{figure}[hc]
\center{\includegraphics[width=162pt]{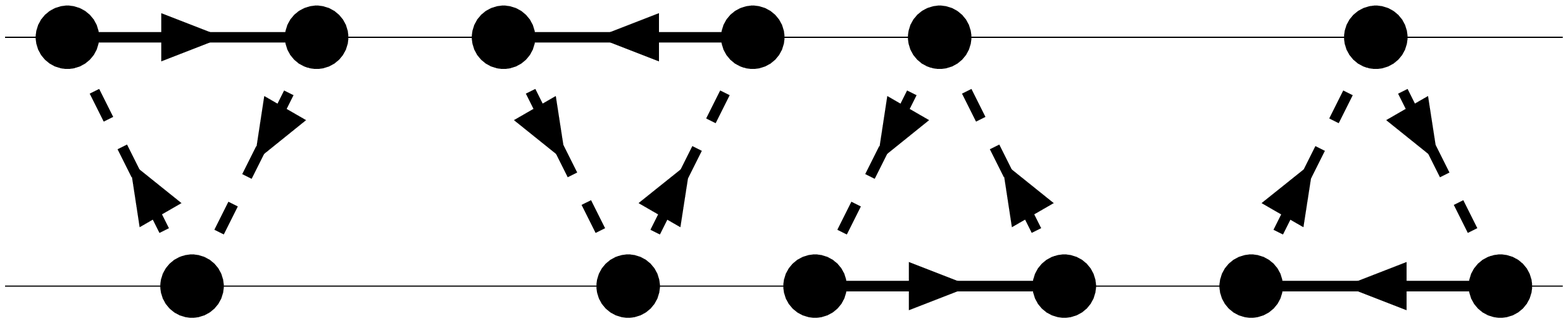}\hspace{1cm}
\includegraphics[width=75pt]{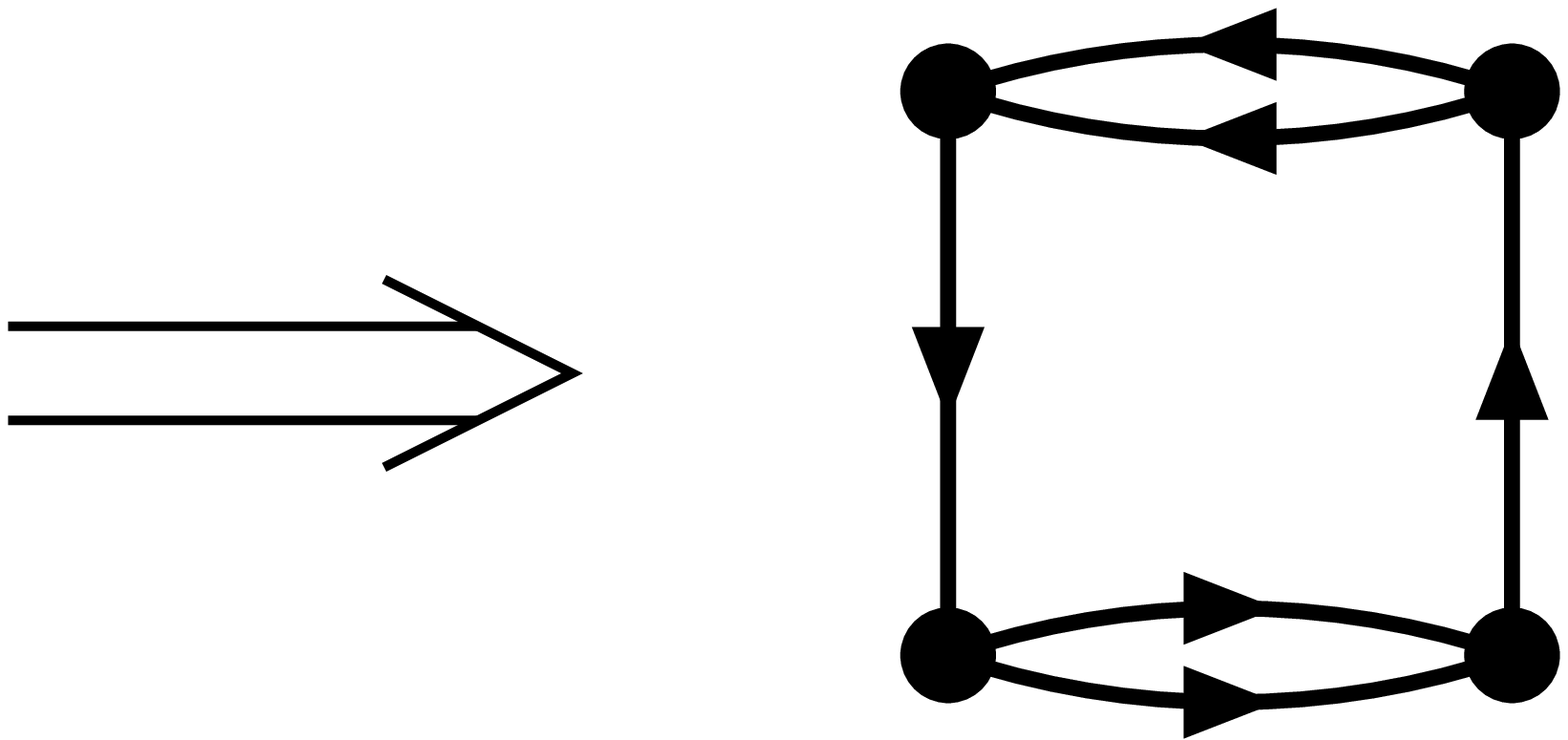}}
\caption{Construction of a graph for the word $1\bar{1}2\bar{2}$ in $SL(3)$. The triangles at left
picture are just cut rhombi from fig.~\ref{fi:lattice}, see also fig.~\ref{fi:treug} in the Appendix~\ref{ap:poisson}. Gluing these triangles, and - for $SL(3)/\Ad H$ - the left and right ends,
one gets the right picture with the exchange matrix constructed from the Cartan matrix \rf{Csl3}.}
\label{fi:todasl3}
\end{figure}
constructed from
glued triangles (see fig.~\ref{fi:todasl3}) and
satisfy the following Poisson bracket relations
\be
\label{pb3}
\{ y_i,x_j\} = C_{ij}y_ix_j,\ \ \ i,j=1,2
\ee
with the Cartan matrix of $\mathfrak{g}=sl_3$
\be
\label{Csl3}
\| C_{ij} \| =\left(
\begin{array}{cc}
  2 & -1 \\
  -1 & 2
\end{array}\right),\ \ \ \ \ \| C^{-1}_{ij} \| =\left(
\begin{array}{cc}
  2/3 & 1/3 \\
  1/3 & 2/3
\end{array}\right)
\ee
To get the integrals of motion,
consider the characteristic polynomial for \rf{gauss3}
\be
\label{Rsl3}
\det \left(\mu + g({\bf x},{\bf y})\right) =
\mu^3 + \mathcal{H}_2({\bf x},{\bf y})\mu^2 + \mathcal{H}_1({\bf x},{\bf y})\mu +1
\ee
where
\be
\label{Rhamsl3}
\mathcal{H}_1({\bf x},{\bf y}) = \Tr\ g^{-1} = \prod_k \left(x_ky_k\right)^{-C^{-1}_{1k}}\cdot\left(1+y_1+y_1x_1+y_1x_1y_2+y_1x_1y_2x_2\right)
\\
\mathcal{H}_2({\bf x},{\bf y}) = \Tr\ g = \prod_k \left(x_ky_k\right)^{-C^{-1}_{2k}}\cdot\left(1+y_2+y_2x_2+y_2x_2y_1+y_2x_2y_1x_1\right)
\ee
become two integrals of motion $\{ \mathcal{H}_1,\mathcal{H}_2\} = 0$ for the Poisson bracket \rf{pb3}.

\subsection{Cluster variables and SL(N) chain}

Explicit formulas can be also written for generic $G=SL(N)$: the product \rf{paramG/AdH} becomes
\be
\label{gaussN}
g({\bf x},{\bf y})
= \underbrace{E_1X_1E_{\bar 1}Y_1}_{g_1}\cdot
\underbrace{E_2X_2E_{\bar 2}Y_2}_{g_2}\cdot\ldots\cdot\underbrace{E_{N-1}X_{N-1}E_{\overline{N-1}}Y_{N-1}}_{g_{N-1}}
\\
Y_i = H_i(y_i)/\det H_i(y_i)^{1/N},\ \ \
X_i = H_i(x_i)/\det H_i(x_i)^{1/N},\ \ \ i=1,\ldots,N-1
\ee
with notations from \rf{genslN},
and the corresponding graph ${ }^{y_1}_{x_1}\Downarrow^{\leftarrow}_{\rightarrow}\Uparrow{ }^{x_2\rightarrow}_{y_2\leftarrow}
\Downarrow^{y_3\leftarrow}_{x_3\rightarrow}\ldots { }^{\rightarrow}_{\leftarrow}\Downarrow^{y_{N-1}}_{x_{N-1}}$ induces the Poisson bracket
\be
\label{pbN}
\{ y_i,x_j\} = C_{ij}y_ix_j,\ \ \ i,j=1,\ldots,N-1
\ee
with the $\mathfrak{g}=sl_N$ Cartan matrix \rf{CslN}. Computing the product in \rf{gaussN} one gets
the Lax operator and its characteristic polynomial
\be
\label{RslN}
\det \left(\mu+g({\bf x},{\bf y})\right) =
\sum_{j=0}^N\mu^j\mathcal{H}_j({\bf x},{\bf y})
\ee
generates $\rank\ SL(N) = N-1$ nontrivial ($\mathcal{H}_0=\mathcal{H}_N=1$ in the accepted normalisation,
and $\mathcal{H}_1 = \Tr\ g^{-1}$, $\mathcal{H}_{N-1} = \Tr\ g$) Poisson-commuting (w.r.t. the bracket \rf{pbN}) integrals of motion $\{ \mathcal{H}_i,\mathcal{H}_j\} = 0$, $i,j=1,\ldots,N-1$, which are
\be
\label{Rhamsl}
\mathcal{H}_j({\bf x},{\bf y}) = \prod_k \left(x_ky_k\right)^{-C^{-1}_{jk}}\cdot Z_j({\bf x},{\bf y})
\ee
where the polynomials
\be
\label{dimer}
Z_j({\bf x},{\bf y}) = \sum_{0\le m_i\le \max(i,N-1-i)}^{m_j\ge m_{j\pm 1}\ge m_{j\pm 2}\ge\ldots}\
\sum_{m_i-1\le n_i\le m_i}\
\prod_i y_i^{m_i}x_i^{n_i}
\ee
have all unit coefficients.

The Darboux co-ordinates are related to the cluster variables by
\be
\label{darbun}
x_i = \exp(-(\alpha_i\cdot q)),
\ \ \ \
y_i = \exp((\alpha_i\cdot P)+(\alpha_i\cdot q)),\ \ \ i=1,\ldots,N-1
\ee
with the ``long momenta'' (cf. with the canonical transformation used in \cite{Faddeev})
\be
\label{Pp}
P = p - {1\over 2}\sum_{k=1}^{N-1}\alpha_k \log\left(1+\exp(\alpha_k\cdot q)\right) =
p + {\d\over\d q}\left( {1\over 2}\sum_{k=1}^{N-1}{\rm Li}_2\left(-\exp(\alpha_k\cdot q)\right)\right)
\ee
where $(N-1)$-vectors $q$, $p$, denote the canonical co-ordinate and momenta in the center of mass frame. Substituting \rf{darbun} into the expression
$\mathcal{H}= \mathcal{H}_1+\mathcal{H}_{N-1}$ one gets the
canonical Toda Hamiltonian \rf{HslN},
where (in the open case in contrast to the periodic chain to be considered below), one has to drop
off the square roots with $\alpha_0$ and $\alpha_N$,
i.e. just to replace $\sqrt{1+\exp(q_0-q_1)}$ and $\sqrt{1+\exp(q_N-q_{N+1})}$ by unities.

\subsection{Loop group $\widehat{SL(N)}$ and $N$-periodic chain
\label{ss:periodic}}

The proposed approach works not only for the finite-dimensional Lie groups \cite{FM11}. It is almost obvious, that to get the periodic Toda chain
one should consider decomposition \rf{paramGloop} for loop groups, containing
extra affine simple root, i.e.  $u=\alpha_0{\bar\alpha}_0\alpha_1{\bar\alpha}_1\ldots\alpha_{N-1}{\bar\alpha}_{N-1}\in \hat{W}\times\hat{W}$. At the level of Lie algebras (see Appendix~\ref{ap:alg}) this gives rise to spectral-parameter dependent Lax matrix in the evaluation representation of the corresponding affine algebra.
For relativistic Toda instead, using cluster formulation of the co-extended loop group $\widehat{SL(N)^\sharp}$, we construct a Poisson submanifold
in $\widehat{SL(N)}/\mbox{Ad}H$, but fixing the spectral parameter becomes a nontrivial issue. It depends
on particular way of locating the shift operators in \rf{paramGloop}: the total
co-extension must be trivial due to \rf{trishift}, and all shift operators can be ``annihilated'', say,
moving all them to the right.

Notice, that moving the shift operator \rf{shiftT} through a spectral parameter dependent
matrix, corresponding to affine root, one multiplies the spectral parameter by dynamical variable. This
results in nontrivial multiplicative renormalisation of the coefficients of the spectral curve equation.
It means in particular, that these coefficients themselves are defined ambiguously and only their
invariant combinations can be the Poisson commuting quantities \cite{FM11,FM12}. However, we propose
here (see also sect.~\ref{ss:dvanadva} below) a localising prescription, which fixes completely this
ambiguity.

For the group $\widehat{SL(N)^\sharp}$ the (normalised to unit determinant)
product \rf{paramG/AdH} for the word
$u= \prod_{j\in\mathbb{Z}_N}\alpha_{j}{\bar\alpha}_{j}$ can be written as
\be
\label{gaussNa}
g(\lambda|{\bf x},{\bf y} )
=\prod_{j\in\mathbb{Z}_N} \mathbf{E}_j\mathbf{H}_j(x_j)
\mathbf{E}_{\bar j}\mathbf{H}_j(y_j) =
\prod_{j\in\mathbb{Z}_N} \mathbf{E}_jX_jT_{x_j}\mathbf{E}_{\bar j}Y_jT_{y_j}=
\\
= \prod_{j\in\mathbb{Z}_N}\mathbf{E}_jX_jT_{x_{\nu_j}}T_{x_{\nu_{j+1}}}^{-1}
\mathbf{E}_{\bar j}Y_jT_{y_{\nu_j}}
T_{y_{\nu_{j+1}}}^{-1}
\ee
Here the new variables $z_{\nu_j}$, $j\in \mathbb{Z}_N$, are introduced by $z_j=z_{\nu_j}/z_{\nu_{j+1}}$ (see definition \rf{numu}). Instead of moving all shift operators to the right, let us choose the following prescription
\be
\label{Aloc}
g(\lambda|{\bf x},{\bf y} )
\simeq \prod_{j\in\mathbb{Z}_N}T_{y_{\nu_j}}^{-1}\mathbf{E}_jX_jT_{x_{\nu_j}}
T_{x_{\nu_{j+1}}}^{-1}\mathbf{E}_{\bar j}Y_jT_{y_{\nu_j}}\simeq
\prod_{j\in\mathbb{Z}_N}T_{x_{\nu_j}}^{-1}T_{y_{\nu_j}}^{-1}\mathbf{E}_jX_jT_{x_{\nu_j}}
\mathbf{E}'_{\bar j}Y_jT_{y_{\nu_j}}
\ee
where $\simeq$ means equality modulo cyclic permutation, and here clearly
\be
\mathbf{E}'_{\bar j} = \mathbf{E}_{\bar j} = E_{\bar j},\ \ \ \ \mathbf{E}_{\bar 0}'=\mathbf{E}_{\bar 0}(\lambda/x_{\nu_{j+1}})
\\
Y_j = H_j(y_j)/\det H_j(y_j)^{1/N},\ \ \
X_j = H_j(x_j)/\det H_j(x_j)^{1/N},\ \ \ j=1,\ldots,N-1
\ee
Now, all factors under the product in the r.h.s. of \rf{Aloc} do not contain shift operators, since
\be
T_{x_{\nu_j}}^{-1}T_{y_{\nu_j}}^{-1}E_jX_jT_{x_{\nu_j}}E_{\bar{j}}Y_jT_{y_{\nu_j}} =
E_jX_jE_{\bar j}Y_j,\ \ \ j=1,\ldots,N-1
\ee
and
\be
\label{g0lam}
T_{x_{\nu_0}}^{-1}T_{y_{\nu_0}}^{-1}\mathbf{E}_0T_{x_{\nu_0}}\mathbf{E}'_{\bar 0}T_{y_{\nu_0}} =
\mathbf{E}_0\left({\lambda \over x_{\nu_0}y_{\nu_0}}\right)\mathbf{E}_{\bar 0}\left({\lambda \over y_{\nu_0}x_{\nu_1}}\right) =
\\
= \mathbf{E}_0\left(\lambda  x_{\mu_{N-1}}y_{\mu_{N-1}}\right)\mathbf{E}_{\bar 0}\left({\lambda y_{\mu_{N-1}}\over x_{\mu_1}}\right)
\equiv {\sf g}_0(\lambda )
\ee
i.e. in the r.h.s. of \rf{Aloc} we got a product of matrices in the evaluation representation of $\widehat{SL(N)}$.
Hence, starting initially with the co-extended loop group $\widehat{SL(N)^\sharp}$, which has
cluster description, we have found a way to construct explicitly the Poisson submanifold in $\widehat{SL(N)}/\mbox{Ad}H$ in terms of the product of the $\widehat{SL(N)}$-valued factors only.
The weight-variables in \rf{g0lam} can be explicitly defined as
\be
x_{\mu_j} = \prod_k x_k^{C^{-1}_{jk}},\ \ \ y_{\mu_j} = \prod_k y_k^{C^{-1}_{jk}},\ \ \ j=1,\ldots,N-1
\ee
with $C_{jk}^{-1}$ being inverse Cartan matrix for $\mathfrak{g}=sl_N$, see Appendix~\ref{ap:promunu}. Finally,
\be
\label{ggM}
g(\lambda|{\bf x},{\bf y} )\simeq{\sf g}_0(\lambda )\cdot g_N({\bf x},{\bf y})
\ee
where the matrix $g_N({\bf x},{\bf y})$, was constructed in \rf{gaussN} for the open chain.

By the standard rules the Poisson brackets, coming from the graph on the torus
\be
\label{pbNp}
\{ y_i, x_j\} = \hat{C}_{ij}y_ix_j,\ \ \ \ \ i,j\in \mathbb{Z}_N
\ee
are defined by (degenerate) Cartan matrix \rf{CslNh} of affine $\mathfrak{g}=\widehat{sl}_N$, with
the Casimir function
\be
\label{casN}
\mathcal{C} =  \prod_{j\in \mathbb{Z}_N} x_j = \prod_{j\in \mathbb{Z}_N} y_j^{-1}
\ee
Fixing the value of the Casimir \rf{casN} one comes to the description of the corresponding Poisson submanifold (of dimension $2(N-1)$) in terms of the non-degenerate Poisson structure \rf{pbN} of the open chain.

\begin{figure}[hc]
\center{\includegraphics[width=200pt]{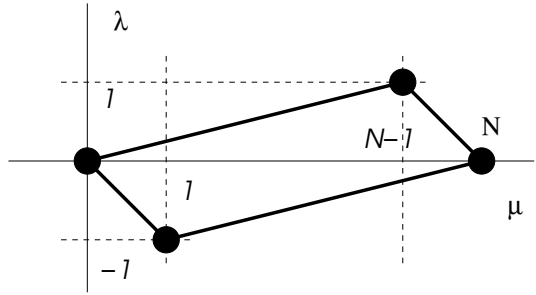}}
\caption{Newton polygon for the $N$-periodic Toda equation \rf{chapo}. The marked boundary
points correspond to the Casimir functions, being just unities in the normalised equation
\rf{chapo}, while the internal points on horisontal axis correspond to the integrals of
motion.}
\label{fi:newtod}
\end{figure}
Vanishing of the characteristic polynomial for \rf{gaussNa} (see fig.~\ref{fi:newtod})
\be
\label{chapo}
\det \left(\mu + g(\lambda|{\bf x},{\bf y} )\right) =
\sum_{j=0}^N \mu^j\mathcal{H}_j({\bf x},{\bf y}) + \mu^{N-1}\lambda +{\mu\over \lambda }
\ee
can be rewritten in standard form of the relativistic Toda spectral curve equation
(see, e.g. \cite{SWcompact})
\be
\label{specu}
w + {1\over w} = P_N(z) = z^{-N/2}\sum_{j=0}^N z^j(-)^{N-j}\mathcal{H}_j({\bf x},{\bf y})
\\
z=-\mu,\ \ \ w=\lambda z^{N/2-1}
\ee
with
two meromorphic differentials ${dz\over z}$ and ${dw\over w}$ having all fixed periods on the curve \rf{specu}.
Formulas \rf{chapo}, \rf{specu} define (in addition to the constant Casimirs  $\mathcal{H}_0=\mathcal{H}_{N-1}=1$) the set of the Poisson commuting integrals of motion
\be
\{ \mathcal{H}_j, \mathcal{H}_k \} = 0,\ \ \ j,k=1,\ldots,N-1
\ee
with respect to the bracket \rf{pbN}, and the canonical Hamiltonian of the integrable system is still given by \rf{HslN}, where the terms with $\alpha_0=\alpha_N$ are no longer dropped off.

The Darboux co-ordinates are introduced again by the transformation
\be
\label{yPp}
x_i = \exp(-(\alpha_i\cdot q)),\ \ \
y_i = \exp(\alpha_i\cdot(P+ q)),\ \ \ \ i\in \mathbb{Z}_N
\ee
valid for the extended set of roots, and now instead of \rf{Pp} one gets
\be
\label{Ppp}
P =  p + {1\over 2}\sum_{k=1}^{N-1}\alpha_k
\log{1+\exp(\alpha_N\cdot q)\over 1+\exp(\alpha_k\cdot q)} =
 p + {\d\over\d q}\left( {1\over 2}\sum_{k=1}^N{\rm Li}_2\left(-\exp(\alpha_k\cdot q)\right)\right)
\ee
where the canonical transformation is again generated by di-logarithm functions.

\paragraph{Example of $\widehat{SL(2)}$.}

\begin{figure}[hc]
\center{\includegraphics[width=162pt]{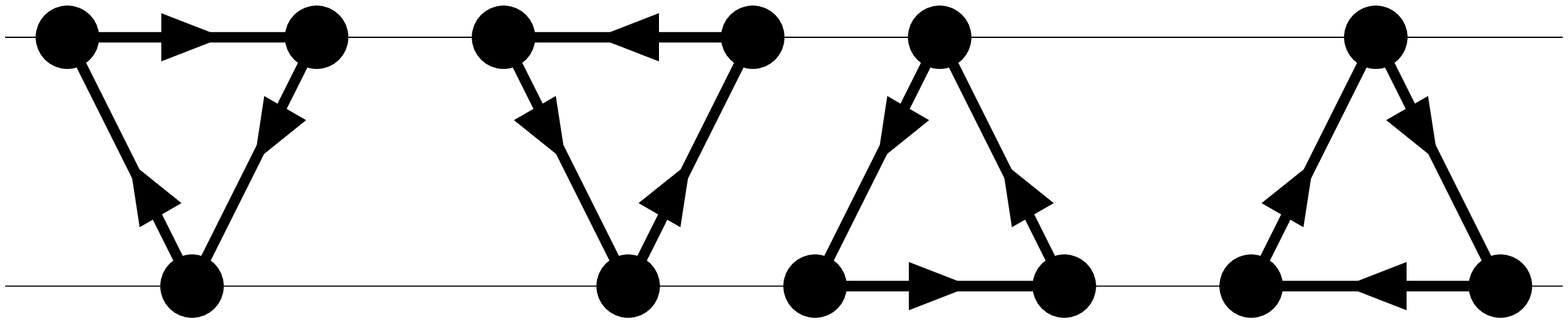}\hspace{1cm}
\includegraphics[width=75pt]{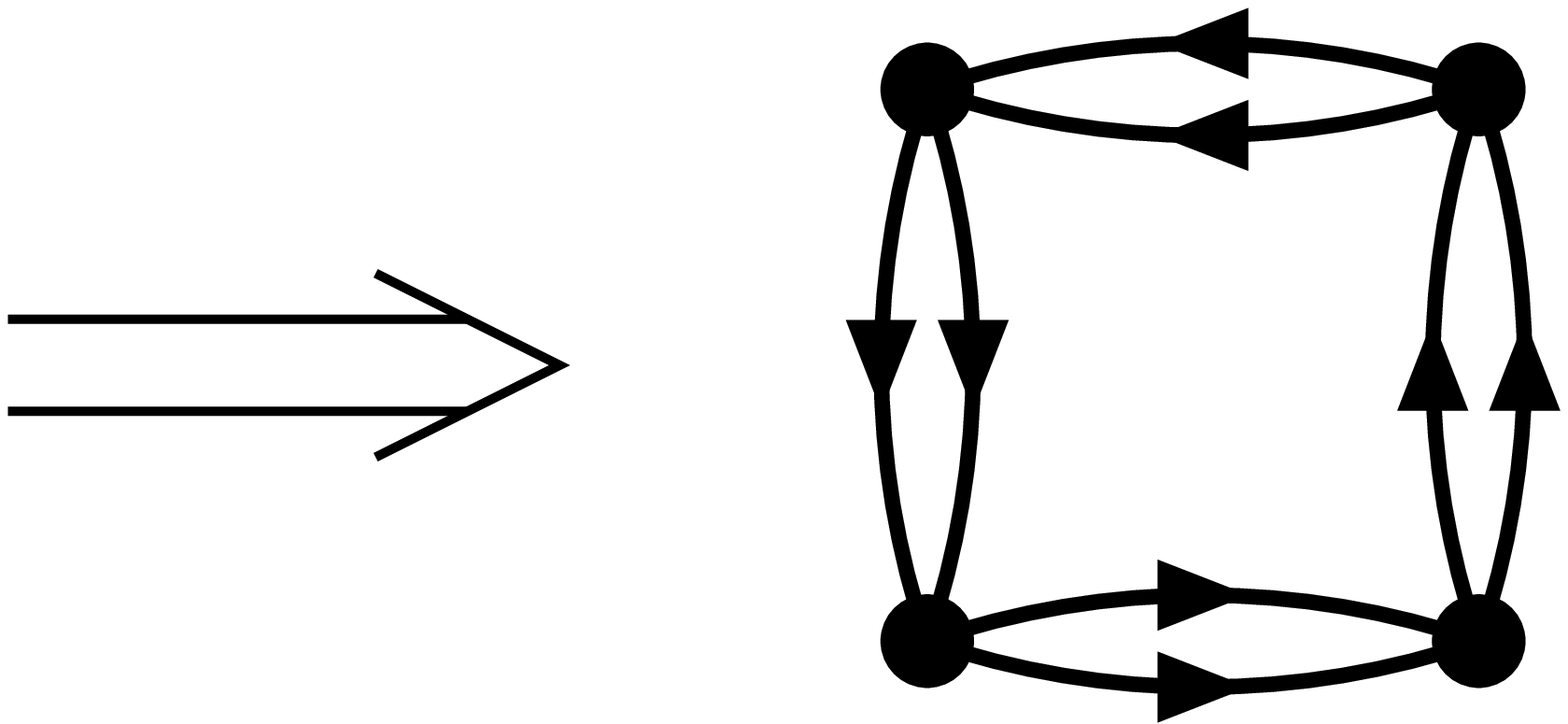}}
\caption{Example of a graph  for the word $0\bar{0}1\bar{1}$ for $\widehat{SL(2)}/\Ad H$. In
contrast to fig.~\ref{fi:todasl3} the arrows, connecting two levels - two simple roots of $\widehat{SL(2)}$ are solid lines, and the resulting - after gluing right picture gets structure of the degenerate Cartan
matrix of $\mathfrak{g}=\widehat{sl_2}$.}
\label{fi:todasl2hat}
\end{figure}
One just takes the map \rf{paramGloop} of the word $0\bar{0}1\bar{1}$ into $\widehat{SL(2})/\Ad\ H$.
According to fig.~\ref{fi:todasl2hat},
the parameterisation can be read from the graph $\Downarrow_{y_0\Rightarrow\ x_0}^{x\ \Leftarrow\ y}\Uparrow$ with four variables in the vertices, satisfying
\be
\label{pb2a}
\{ y,x\} = 2yx
\ee
as in the ``open'' $SL(2)$-case, completed now by
\be
\label{pb2aa}
\{ y_0,x\} = -2y_0x, \ \ \ \{ y,x_0\} = -2yx_0, \ \ \ \ \{ y,y_0\} = \{ x,x_0\} = 0
\ee
The structure of the Poisson brackets \rf{pb2a}, \rf{pb2aa} is encoded in the degenerate Cartan matrix \rf{CslNh}
of $\mathfrak{g}=\widehat{sl}_2$, and product ${\cal C} = xx_0= {1\over yy_0}$ is the Casimir function.
The only nontrivial integral of motion in this example
\be
\label{HA}
\mathcal{H}= 
{1+y+xy+{\mathcal C}^{-1}x\over \sqrt{xy}} =
\left(e^{p}+e^{-p}\right)\sqrt{1+\exp(q)}\sqrt{1+{\mathcal C}^{-1}\exp(-q)}
\ee
is the Hamiltonian of the periodic 2-particle chain \rf{HslN} upon
\be
\label{dar2a}
x=\exp(-q),\ \ \ y=\exp(2p+q){1+{\mathcal C}^{-1}e^{-q}\over 1+ e^q}
\ee

\paragraph{Example of $\widehat{SL(3)}$.}
In this case the characteristic polynomial \rf{chapo}, or the spectral curve equation \rf{specu}
gives two nontrivial integrals of motion
\be
\label{Rhamsl3p}
\mathcal{H}_1 = \prod_k \left(x_ky_k\right)^{-C^{-1}_{1k}}\cdot\left(1+y_1+y_1x_1+y_1x_1y_2+y_1x_1y_2x_2
+{\cal C}^{-1}x_1x_2\right)
\\
\mathcal{H}_2 = \prod_k \left(x_ky_k\right)^{-C^{-1}_{2k}}\cdot\left(1+y_2+y_2x_2+y_2x_2y_1+y_2x_2y_1x_1+
{\cal C}^{-1}x_1x_2\right)
\ee
Only the last terms in the r.h.s.'s of \rf{Rhamsl3p} (expressed via the Casimir ${\cal C}=x_0x_1x_2 = {1\over y_0y_1y_2}$) differ them from the Hamiltonians of non-periodic chain \rf{Rhamsl3}.

\subsection{Mutations and discrete flows}

There are the \emph{discrete} transformations, leaving invariant the Toda Hamiltonians, constructed as a sequence of mutations (see \rf{mutz} in Appendix~\ref{ap:poisson}) of the corresponding graphs. For \rf{Rhamsl}, \rf{dimer} these
\be
\label{disl}
\tilde{y}_i^{-1} = y_i\prod_j\left(1+x_j^{{\rm sgn}(C_{ij})}\right)^{-C_{ij}},\ \ \ \
\tilde{x}_i = x_i^{-1}\prod_j\left(1+\tilde{y}_j^{-{\rm sgn}(C_{ij})}\right)^{C_{ij}}
\ee
come from mutations of the graph ${ }^{y_1}_{x_1}\Downarrow^{\leftarrow}_{\rightarrow}\Uparrow{ }^{x_2\rightarrow}_{y_2\leftarrow}
\Downarrow^{y_3\leftarrow}_{x_3\rightarrow}\ldots { }^{\rightarrow}_{\leftarrow}\Downarrow^{y_{N-1}}_{x_{N-1}}$. For example, the Hamiltonians \rf{Rhamsl3} are invariant under the transformations\footnote{In the dual variables $x_j=\prod_k b_k^{C_{jk}}$ and $y_j=\prod_k a_k^{-C_{jk}}$
they acquire the form of the discrete bilinear Hirota equations
\be
\label{hirsl3}
a_i{\tilde a}_i = b_i^2 + b_{i+1}b_{i-1}
\\
b_i{\tilde b}_i = {\tilde a}_i^2 + {\tilde a}_{i+1}{\tilde a}_{i-1}
\ee
with, e.g. for $SL(3)$ $b_0=b_3=a_0=a_3=2$.}
\be
\label{disl3}
\tilde{y}_1 = {1\over y_1}\left(1+x_1\right)^{-2}\left(1+{1\over x_2}\right),\ \ \
\tilde{y}_2 = {1\over y_2}\left(1+x_2\right)^{-2}\left(1+{1\over x_1}\right)
\\
\tilde{x}_1 = {1\over x_1}\left(1+{1\over \tilde{y}_1}\right)^{2}\left(1+\tilde{y}_2\right)^{-1},\ \ \
\tilde{x}_2 = {1\over x_2}\left(1+{1\over \tilde{y}_2}\right)^{2}\left(1+\tilde{y}_1\right)^{-1}
\ee
leaving invariant the graph ${ }^{x_1}_{y_2}\downarrow^{\Leftarrow}_{\Rightarrow}\uparrow{ }^{y_1}_{x_2}$.
The affine Toda Hamiltonians \rf{chapo}, \rf{specu} are invariant under the transformations
\be
\label{dislaff}
\tilde{y}_i^{-1} = y_i\prod_j\left(1+x_j^{{\rm sgn}(\hat{C}_{ij})}\right)^{-\hat{C}_{ij}},\ \ \ \
\tilde{x}_i = x_i^{-1}\prod_j\left(1+\tilde{y}_j^{-{\rm sgn}(\hat{C}_{ij})}\right)^{\hat{C}_{ij}}
\ee
coming from similar mutations of the graph with the exchange matrix constructed from $\hat{C}_{ij}$ \rf{CslNh}.

These transformations can be considered as discrete flows in spirit of \cite{HKKR}. For example, in the decomposition \rf{gxy} into the product of
upper-triangular and lower-triangular matrix change the order, i.e.
\be
\label{mutdec}
g(x,y) \simeq g_+(x,y)\cdot g_-(x,y) \simeq H(y)EH(x)F =
\\
= H(y)H(1+x)FH(x^{-1})EH(1+x) \simeq H\left(y(1+x)^2\right)FH\left(x^{-1}\right)E\simeq
\\
\simeq \tilde{g}_-\left( \tilde{x}, \tilde{y}\right)\cdot \tilde{g}_+\left( \tilde{x}, \tilde{y}\right)
\ee
where we have used the commutation relations from \cite{Drinfeld}. The decomposition in the r.h.s. in ``reverse order'' corresponds to changing
co-ordinates by sequence of mutations
\be
\label{mut}
\left( \tilde{y}, \tilde{x}\right)  = \mu_{y(1+x)^2}\left(y(1+x)^2,x^{-1}\right)\circ\mu_x (y,x)=
\left(y^{-1}(1+x)^{-2}, x^{-1}(1+y(1+x)^2)^2\right)
\ee
for the graph $x\Leftarrow y$.
Note, that we consider the product in \rf{mutdec} modulo cyclic permutations and conjugation
by a Cartan element. The same is certainly true for the ``reflected composition'', when we first
make mutation in $y$- and then in $x$-vertices.

\setcounter{equation}0
\section{More integrable models
\label{ss:nN}}

We have already pointed out that loop groups allow to construct much wider class of integrable systems
\cite{FM11,FM12}. We shall demonstrate it here on few examples, starting from the well-known ``dual''
representation for the Toda chains.

\subsection{$2\times 2$ formalism for relativistic Toda
\label{ss:dvanadva}}

\begin{figure}[tb]
\epsfysize=3.5cm
\centerline{\epsfbox{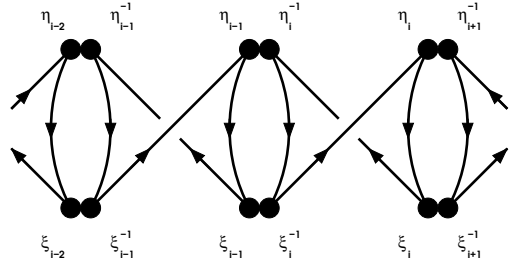}}
\caption{\sl The product of $\Xi$-operators turns into the product of $L$-matrices
in \rf{prMNL} after change of the variables.}
\label{fi:xe22}
\end{figure}

\begin{figure}[tb]
\epsfysize=3cm
\centerline{\epsfbox{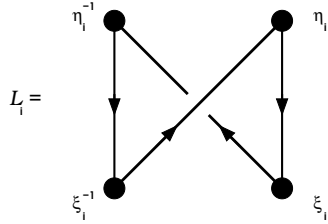}}
\caption{\sl Representation of the $L$-matrix
from \rf{L22}, which is a Cartan conjugation of \rf{Phim} in the form of the twisted oriented plaquette.}
\label{fi:Lxe}
\end{figure}

The same graph $\Downarrow_{y_0\ \Rightarrow x_0}^{x\ \Leftarrow\ y}\Uparrow$ for the $\widehat{SL(2)}$ Toda system (when rotated clockwise, i.e.
$\Downarrow_{x_0\ \Rightarrow\ y}^{y_0\ \Leftarrow\ x}\Uparrow\ ={}_{x_0}\downarrow_{\rightarrow}^{ \leftarrow}\uparrow_y^x \uparrow_{\leftarrow}^{\rightarrow}\downarrow^{y_0}$) can be associated with another decomposition
\be
\label{sl222}
\mathbf{H}_0(x_0)\mathbf{E}_0\mathbf{E}_{\bar 1}\mathbf{H}_0(y)\mathbf{H}_1(x)\mathbf{E}_{\bar 0}\mathbf{E}_1\mathbf{H}_1(y_0)\ \simeq\
\mathbf{H}_0(x_0)\mathbf{E}_0\omega \mathbf{E}_{\bar 0}\mathbf{H}_1(y)\cdot
\mathbf{H}_0(x)\mathbf{E}_0\omega \mathbf{E}_{\bar 0}\mathbf{H}_1(y_0) \equiv
\\
\equiv \Xi(x_0,y)\cdot\Xi(x,y_0)
\ee
where $\simeq$ means as usually equality modulo cyclic permutation. Here
\be
\label{xiop}
\Xi(x,y) = \mathbf{H}_0(x)\mathbf{E}_0\omega \mathbf{E}_{\bar 0}\mathbf{H}_1(y) =
\\
= T_x\cdot \left(
\begin{array}{cc}
   1 & 0 \\
 \lambda  & 1
\end{array}\right)\left(
\begin{array}{cc}
   0 & \lambda^{-1/2} \\
 \lambda^{1/2} & 0
\end{array}\right)\left(
\begin{array}{cc}
   1 & 1/\lambda \\
   0 & 1
\end{array}\right)\left(
\begin{array}{cc}
  y^{1/2} & 0 \\
  0 & y^{-1/2}
\end{array}\right)\cdot T_y =
\\
= T_x\cdot \Phi(\lambda)\cdot YT_{y}
\ee
with
\be
\label{Phim}
\Phi(\lambda ) = \mathbf{E}_0\omega \mathbf{F}_0 = \left(
\begin{array}{cc}
   0 & 1/\sqrt{\lambda } \\
 \sqrt{\lambda } & \sqrt{\lambda }+1/\sqrt{\lambda }
\end{array}\right)
\\
\omega = \sqrt{\lambda}\Lambda = \left(
\begin{array}{cc}
   0 & 1/\sqrt{\lambda} \\
 \sqrt{\lambda} & 0
\end{array}\right),\ \ \ \
\omega^2 = {\bf 1}
\ee
\begin{figure}[tb]
\epsfysize=3.5cm
\centerline{\epsfbox{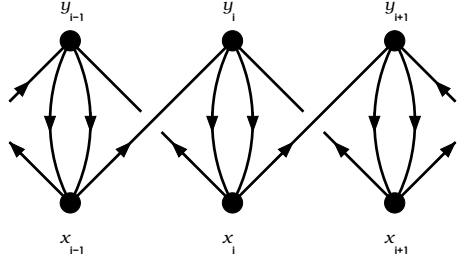}}
\caption{\sl Diagrammatic representation of the product of $\Xi$-operators 
in \rf{prMN}.}
\label{fi:yx22}
\end{figure}
For the $\widehat{SL(3)}$ system, corresponding to the graph $\Downarrow_{x_0\ \rightarrow\ y_1}^{y_0\ \leftarrow\ x_1}\Uparrow_{\leftarrow\ x_2}^{\rightarrow\ y_2}\Downarrow_{\rightarrow\ y_0}^{\leftarrow x_0}$ (i.e. glued with a vertical twist), one has to consider instead of \rf{sl222} the 3-product
$\Xi(x_0,y_1)\Xi(x_1,y_2)\Xi(x_2,y_0)$, which is just a particular case of generic $\widehat{SL(N)}$ expression
\be
\label{prMN}
\Xi(x_0,y_1)\Xi(x_1,y_2)\ldots\Xi(x_{N-1},y_N)
\ee
It is more convenient to rewrite \rf{prMN}, in new variables
\be
\label{xyxe}
x_i = {\xi_i\over\xi_{i+1}},\ \ \ y_i = {\eta_i\over\eta_{i+1}},\ \ \ i=1,\ldots,N-1
\\
\xi_N = \xi_0,\ \ \ \eta_N = \eta_0
\ee
so that the Poisson brackets \rf{pbN} are induced by
\be
\label{xietaN}
\{ \eta_i,\xi_j\} = \delta_{ij}\eta_i\xi_j,\ \ \ \ i,j=1,\ldots,N
\ee
and not to impose any restriction onto the products $\prod_{j=1}^N\xi_j$ and
$\prod_{j=1}^N\eta_j$ (corresponding to the $GL(N)$ instead of the $SL(N)$ group, relating dynamic variables to the basis vectors of $N$-dimensional space, see notations in Appendix~\ref{ap:promunu}).
Similar to \rf{yPp}, the co-ordinates and momenta $\{ q_i,p_j\} = \delta_{ij}$ for \rf{xietaN} are introduced by
\be
\label{xew}
\xi_i = \exp(-q_i),\ \ \
\eta_i = \exp(P_i+q_i),\ \ \ i=1,\ldots,N
\\
P_i=p_i+{\d\over\d q_i}\left( {1\over 2}\sum_{k=1}^N
{\rm Li}_2\left(-\exp(q_k-q_{k+1})\right)\right)
\ee
The product of $\Xi$-operators in \rf{prMN}, again up to cyclic permutation,
can be rewritten as
\be
\label{prMNL}
\Xi(x_0,y_1)\Xi(x_1,y_2)\ldots\Xi(x_{N-1},y_N) \simeq \prod_{j=1}^N L(\eta_j,\xi_j;\lambda )
= {\cal T}_N(\lambda )
\ee
since (see fig.~\ref{fi:xe22})
\be
\label{XiL}
\ldots\Xi(x_{i-1},y_i)\ldots = \ldots T_{x_{i-1}}\Phi(\lambda )\left(\begin{array}{cc}
                                                                y_i^{1/2} & 0 \\
                                                                0 & y_i^{-1/2}
                                                              \end{array}
\right)T_{y_i}\ldots =
\\
\stackreb{\rf{xyxe}}{=}\ \ldots T_{\xi_{i-1}}T_{\xi_i}^{-1}\Phi(\lambda )\left(\begin{array}{cc}
                                                                \eta_i^{1/2} & 0 \\
                                                                0 & \eta_i^{-1/2}
                                                              \end{array}
\right)\left(\begin{array}{cc}
                                                                \eta_{i+1}^{-1/2} & 0 \\
                                                                0 & \eta_{i+1}^{1/2}
                                                              \end{array}
\right)T_{\eta_i}T_{\eta_{i+1}}^{-1}\ldots =
\\
= \ldots \left(\begin{array}{cc}
                                                                \eta_i^{-1/2} & 0 \\
                                                                0 & \eta_i^{1/2}
                                                              \end{array}
\right)T_{\eta_i}^{-1}T_{\xi_i}^{-1}\Phi(\lambda )\left(\begin{array}{cc}
                                                                \eta_i^{1/2} & 0 \\
                                                                0 & \eta_i^{-1/2}
                                                              \end{array}
\right)T_{\eta_i}T_{\xi_i}\ldots
\ee
where in the r.h.s. we have collected only the factors, corresponding to the $i$-th cite.
It is clear from \rf{XiL}, that in the r.h.s. of \rf{prMNL} one gets the
product of the matrices (cf. with \cite{FT,Suris})
\be
\label{L22}
L_j(\lambda ) =
\left(\mathbf{H}_1(\eta_j)\mathbf{H}_0(\xi_j)\right)^{-1}\cdot\Phi(\lambda )\cdot\mathbf{H}_1(\eta_j)\mathbf{H}_0(\xi_j) =
\\
= \left(\begin{array}{cc}
                      0 & \sqrt{\xi_j\over \lambda \eta_j} \\
                      \sqrt{\lambda \eta_j\over \xi_j}\ \ & \ \sqrt{\lambda \over \eta_j\xi_j}+\sqrt{\eta_j\xi_j\over \lambda }
                    \end{array}\right)=
\left(\begin{array}{cc}
                      0 & {e^{-P_j/2-q_j}\over \sqrt{\lambda }} \\
                      \sqrt{\lambda }e^{P_j/2+q_j}\ \ & \ \sqrt{\lambda }e^{-P_j/2}+{e^{P_j/2}\over \sqrt{\lambda }}
                    \end{array}\right)
                    \\
j=1,\ldots,N
\ee
The dependence on dynamic variables arises by the Cartan conjugation of the matrices \rf{Phim} in
evaluation representation of $\widehat{SL(2)}$, see fig.~\ref{fi:Lxe}, and these
$L$-operators obey the $r$-matrix Poisson relations
\be
\label{PbS}
\left\{ L_i(\lambda ) \stackreb{,}{\otimes}L_j(\lambda ') \right\}= -\half\delta_{ij}\left[r_{\rm trig}\left(\lambda ,\lambda '\right), L_i(\lambda ) \stackreb{,}{\otimes}L_j(\lambda ')\right]
\ee
with the trigonometric classical $r$-matrix \rf{rtrig} (see Appendix~\ref{ap:rmatr}). The same commutation relation obviously holds for the monodromy matrices \rf{prMNL}
\be
\label{PbT}
\left\{ {\cal T}_N(\lambda ) \stackreb{,}{\otimes}{\cal T}_N(\lambda ') \right\}= -\half\left[r_{\rm trig}\left(\lambda ,\lambda '\right), {\cal T}_N(\lambda ) \stackreb{,}{\otimes}{\cal T}_N(\lambda ')\right]
\ee
Therefore, the traces ${\sf T}_N(\lambda ) = \Tr\ {\cal T}_N(\lambda )$, or the coefficients of
the characteristic polynomial for the monodromy matrix
\be
\label{chpN}
\det ({\cal T}_N(\lambda )+\mu) = \mu^2 + {\sf T}_N(\lambda )\mu + 1
\ee
Poisson-commute $\{ {\sf T}_N(\lambda ) ,{\sf T}_N(\lambda ') \}=0$ w.r.t. \rf{xietaN}, and produce
$\lambda ^{N/2}{\sf T}_N(\lambda ) = \sum_{j=0}^N\lambda ^j\tilde{R}_j(\bxi,\bbeta)$ the family of $N$ independent integrals of motion:
\be
\label{RR}
\tilde{R}_0(\bxi,\bbeta) = \prod_{k=1}^N\sqrt{\eta_k\xi_k} = {1\over \tilde{R}_N(\bxi,\bbeta)}
\\
\tilde{R}_j(\bxi,\bbeta) = \tilde{R}_0(\bxi,\bbeta)^{1-{2j\over N}}
\mathcal{H}_j({\bf x},{\bf y}),\ \ \ \ j=1,\ldots,N-1
\ee
where in the r.h.s. one finds the Hamiltonians \rf{specu}, as functions of the cluster
variables \rf{xyxe}.

\subsection{$n\times N$ models from square lattice}

The $2\times 2$ formalism for relativistic Toda of the previous section is based on construction
of a Poisson submanifold in $\widehat{SL(2)}$, corresponding to the word of length $2N$, and it has an obvious $n\times N$ generalisation to other co-extended loop groups.
\begin{figure}[hc]
\center{\includegraphics[width=130pt]{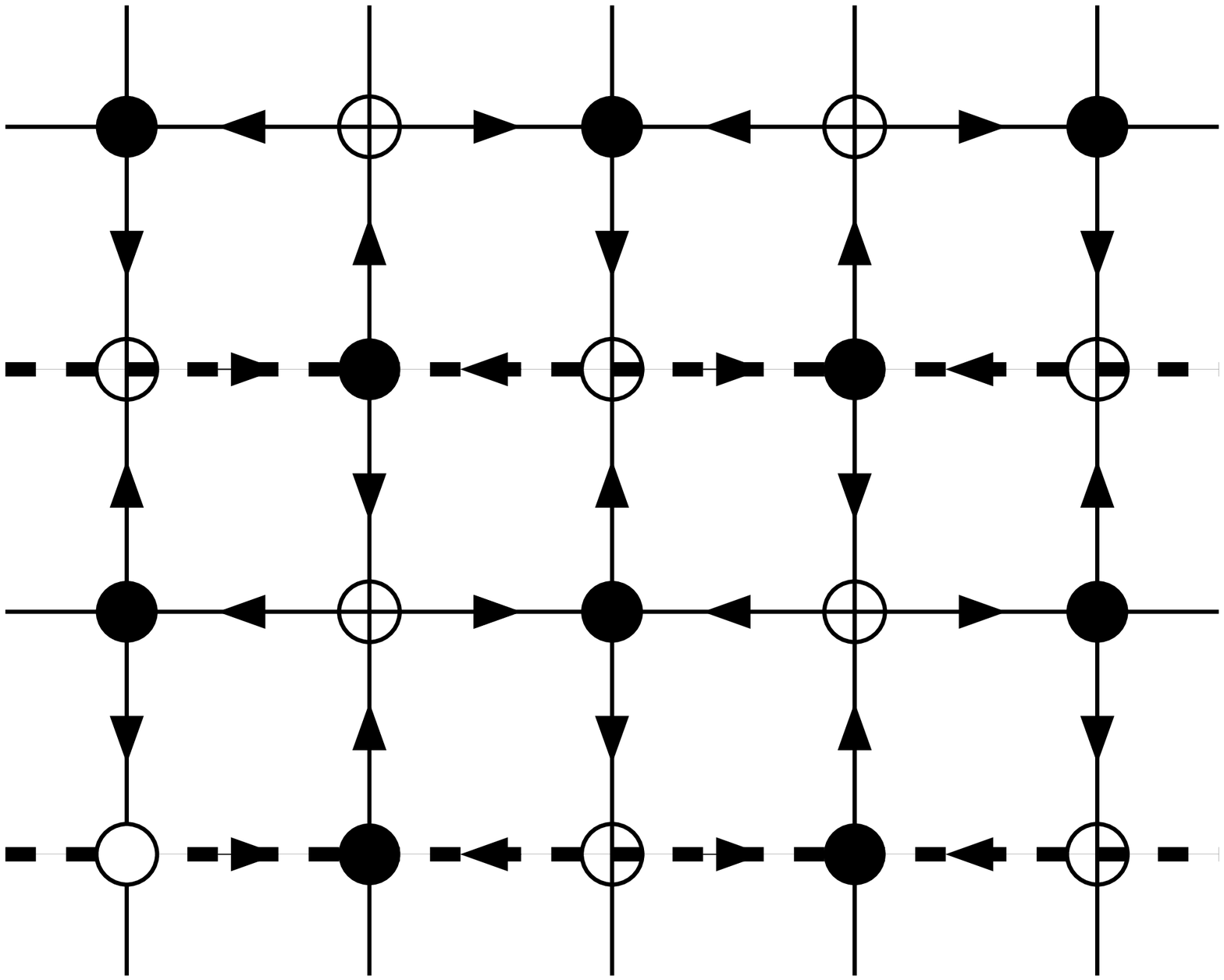} \qquad \qquad \qquad
\includegraphics[width=140pt]{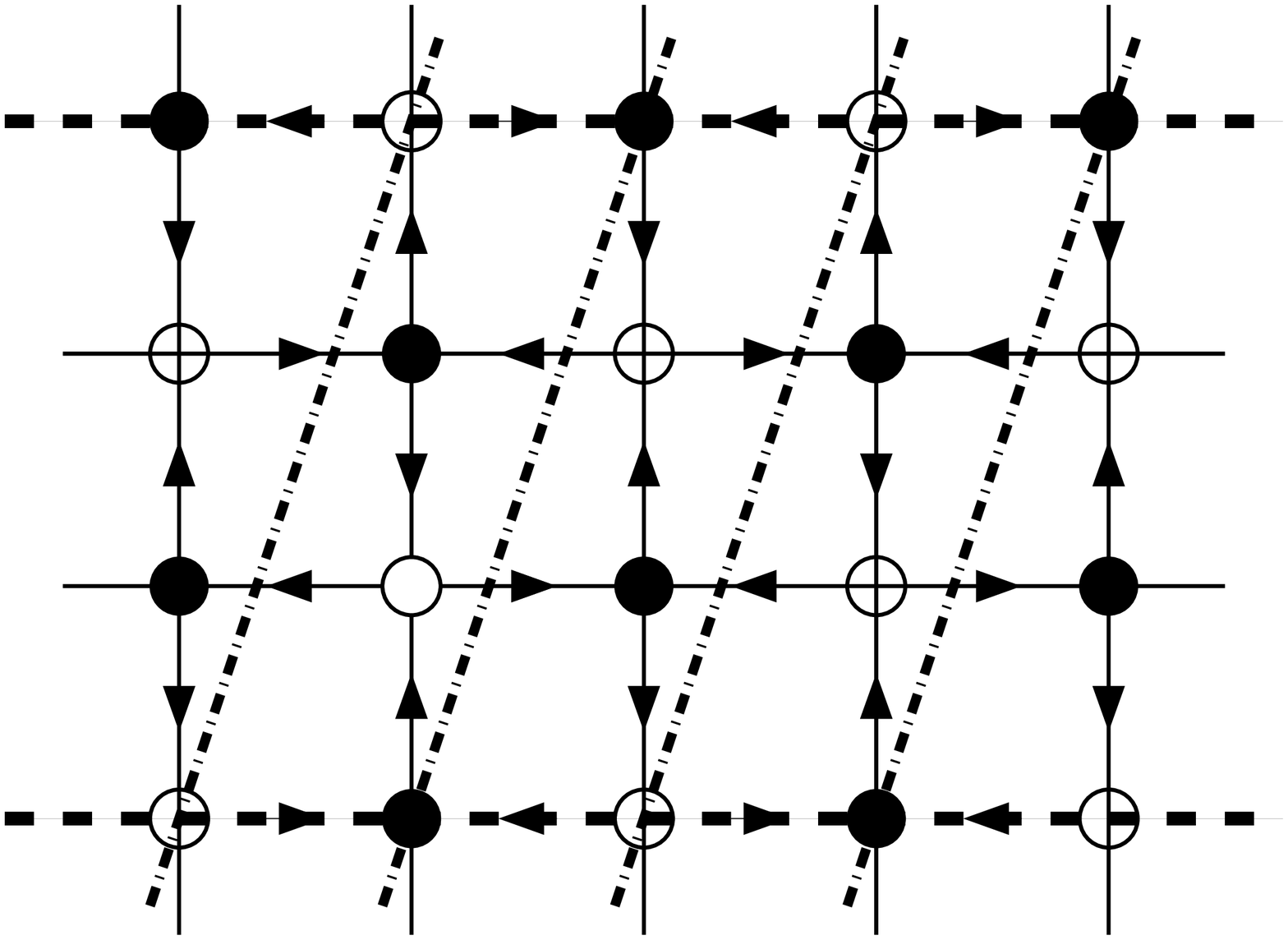}
 }
\caption{Contraction of an infinite square lattice by imposing 2-periodicity constraint in vertical direction (left picture), resulting in the graph $\cdots{}^\rightarrow_\leftarrow\Downarrow_{x_0\ \rightarrow\ y_1}^{y_0\ \leftarrow\ x_1}\Uparrow_{\leftarrow\ x_2}^{\rightarrow\ y_2}\Downarrow\cdots\Downarrow_{\rightarrow\ y_N}^{\leftarrow x_N}\Uparrow^\rightarrow_\leftarrow\cdots$. Another contraction (right picture) of an infinite square lattice by imposing 3-periodicity constraint in vertical direction, together with an extra horisontal 1-step shift.}
\label{fi:23pred}
\end{figure}
A subclass of these systems can be constructed from gluing of the infinite-dimensional square lattice
\cite{FockP}, see fig.~\ref{fi:23pred}. For example, identifying the horisontal lines in such lattice by $2$-shift as on left fig.~\ref{fi:23pred}, one gets the graph,
consisting of two horisontal sets of vertices, connected by vertical double-edges. After imposing
an extra $N$-periodicity in horisontal direction (for $N\geq 2$, with an extra twist for the odd $N$) we recover in this way construction of monodromy matrix \rf{prMNL} for affine Toda chain.
The Poisson
stricture \rf{pbNp} is then induced by the simple Poisson structure on the lattice,
determined by single arrows between neighbor (black and white) vertices, drawn in the way that each plaquette of the lattice is oriented.

Some generalizations of the Toda systems naturally arise for different gluing of the same lattice.
For example, consider now 3-periodicity (instead of 2-periodicity), with
an extra 1-step horisontal shift to be consistent with the orientation of arrows, see right fig.~\ref{fi:23pred}.
Such 3-reduced lattice can be further used for construction of the monodromy matrix in terms of the co-extended $\widehat{SL^\sharp(3)}$ loop group, and we present this construction in next section (of course, this is easily generalized to the loop groups $\widehat{SL^\sharp(n)}$ for any $n\geq 2$).

The Poisson structure given by gluing as at right fig.~\ref{fi:23pred} can be written as
\be
\label{uvbra3}
\{y_j,x_i\}=y_jx_i\ \ \mbox{ if } j=i \mbox{ or } j=i+3,\\
\{y_j,x_i\}=-y_jx_i \mbox{ if } j=i+1 \mbox{ or } j=i+2
\ee
for some labeling by $y$'s the white vertices, and by $x$'s - the black ones,
while all the other brackets vanish. Another useful representation for the same Poisson structure
\be
\label{schwartz}
\{ Y_k,Y_{k\pm 1}\} = \mp Y_kY_{k\pm 1}
\\
\{ X_k,X_{k\pm 1}\} = \pm X_kX_{k\pm 1}
\ee
comes after the substitution
\be
\label{glick}
x_i = Y_iX_{i+1},\quad y_i=(X_{i-1}Y_{i-1})^{-1}
\ee
The variables in \rf{uvbra3}, \rf{schwartz} were already introduced in the context of integrable system,
related to the pentagram map \cite{OT,Glick}, and we demonstrate now, that this is a generalized Toda system in the sense of our approach  \cite{FM11,FM12}.

\subsection{Integrable system for $n=3$}

For $n=3$, similarly to \rf{xiop}, one starts with the product \cite{FockP}
\be
\label{xiopp3}
\Xi(y,x)= \mathbf{H}_0(y)\mathbf{E}_0 \Lambda \mathbf{E}_{\bar 0} \mathbf{H}_{2}(x)
\ee
of the elements of $\widehat{SL^\sharp(3)}$, where we use the notations \rf{notloop}, \rf{loopextra}, \rf{lpshft}. These generators are labeled
modulo 3, e.g. $\mathbf{H}_{-1}(z)=\mathbf{H}_2(z)$ etc, while the co-ordinates will be considered further as $N$-periodic, i.e. $x_{i+N}=x_i$, $y_{i+N}=y_i$, for an arbitrary $N\geq 3$.

The product $\Xi(y_i,x_i)\cdots \Xi(y_j,x_j)$ can be transformed by taking all the automorphisms $\Lambda$ to the right:
\be
\label{xiprod}
\Xi(y_i,x_i)\ldots \Xi(y_j,x_j) =\mathbf H_0(y_i)\mathbf E_0 \Lambda \mathbf{E}_{\bar 0}  \mathbf H_{-1}(x_i)\ldots\mathbf H_0(y_j)\mathbf E_0 \Lambda \mathbf{E}_{\bar 0} \mathbf H_{-1}(x_j)=
\\
=
{\mathbf H}_0(y_i){\mathbf E}_0 {\mathbf E}_{\bar 1} {\mathbf H}_0(x_i)
{\mathbf H}_1(y_{i+1}){\mathbf E}_1 {\mathbf E}_{\bar 2} {\mathbf H}_1(x_{i+1})\ldots
\\
\ldots{\mathbf H}_{j-i-1}(y_j){\mathbf E}_{j-i-1}{\mathbf E}_{\overline{j-i}}{\mathbf H}_{j-i-1}(x_j)\Lambda^{j-i+1}
\ee
Therefore, up to the power of $\Lambda$ at the end, this expression particular case of the map \rf{paramG/AdH} (see also \cite{Drinfeld}). It corresponds to the word $u=\alpha_0\bar{\alpha}_1\alpha_1\bar{\alpha}_2\cdots \bar{\alpha}_{j-i}$, and the corresponding graph
is equivalent to the right fig.~\ref{fi:23pred}, giving rise exactly to the Poisson bracket \rf{uvbra3}.

\begin{figure}[tb]
\epsfysize=4.5cm
\centerline{\epsfbox{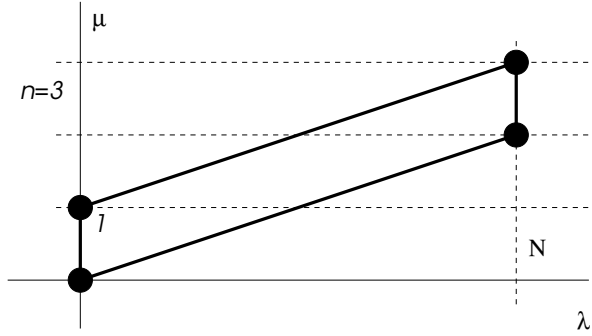}}
\caption{\sl The Newton polygon for the curve \rf{chp}, which leads to the genus formula \rf{genus}.}
\label{fi:newtonOT}
\end{figure}

Consider now a generic factor in the
product \rf{xiopp3}, using co-ordinates \rf{glick}
\be
\label{loc}
\ldots \Xi(y_k,x_k;\lambda)\ldots\ = \ldots \mathbf H_0(y_k)\mathbf E_0 \Lambda \mathbf F_0 \mathbf H_{2}(x_k)\ldots =
\\
 \stackreb{\rf{glick}}{=}\ \ldots H_2(Y_{k-1}X_k)T_{Y_{k-1}X_k}T^{-1}_{X_{k-1}Y_{k-1}} \Phi(\lambda)
 H_2(Y_kX_{k+1})T_{Y_kX_{k+1}}T^{-1}_{X_kY_k}\ldots
\ee
where
\be
\label{fi3}
\Phi(\lambda) = \mathbf E_0 \Lambda \mathbf F_0
 =\begin{pmatrix}
0&0&\lambda^{-1}\\
1&0&\lambda^{-1}\\
0&1&1
  \end{pmatrix}
\ee
and in the r.h.s. or \rf{loc} we have presented explicitly all operators, depending on the variables $\{ Y_k,X_k\}$ at $k$-th site. This can be further rewritten as
\be
\label{prloc3}
\ldots\ \underbrace{H_2(X_k)T_{X_k}\Phi(\lambda)T^{-1}_{X_k} H_2(Y_k)}_k\
\underbrace{H_2(X_{k+1})\ldots }_{k+1}\ \ldots =
\\
\ldots\ \underbrace{H_2(X_k)\Phi(\lambda X_k) H_2(Y_k)}_k\
\underbrace{H_2(X_{k+1})\ldots }_{k+1}\ \ldots
\ee
i.e. as a product of the matrices (instead of original operators \rf{xiopp3} from $\widehat{PGL^\sharp(3)}$)
\be
\label{lax33}
L_k(\lambda) = L(\lambda;X_k,Y_k)= H_2(X_k)\Phi(\lambda X_k) H_2(Y_k) =
 \begin{pmatrix}
0&0&1/\lambda\\
X_kY_k&0&1/\lambda\\
0&Y_k&1
  \end{pmatrix}
\ee
 each corresponding to
particular $k$-th site. Using \rf{loc}, \rf{prloc3} the initial formula \rf{xiprod}, up cyclic permutations,  is equivalent to the
product
\be
\label{monxy}
\Xi(y_0,x_0)\ldots \Xi(y_{N-1},x_{N-1}) \simeq {\cal T}_N(\lambda) = L_0(\lambda)\ldots L_{N-1}(\lambda)
\ee
The characteristic polynomial for \rf{monxy}
\be
\label{chp}
\lambda^N\det\left({\cal T}_N(\lambda)+\mu\right) = \mu^3\lambda^N + \mu^2C_2(\lambda)+\mu C_1(\lambda)+1
\ee
gives the spectral curve equation. The determinant of \rf{monxy}
\be
\label{detT}
\lambda^N\det {\cal T}_N(\lambda) = \lambda^N\prod_{j=0}^N \det L_j(\lambda) = \prod_{j=0}^N X_j \left(\prod_{j=0}^N Y_j\right)^2 =1
\ee
is proportional to a Casimir function for \rf{schwartz} and fixed to be unity in \rf{chp}.
The coefficients of
the polynomials $C_i(\lambda)$, $i=1,2$ in \rf{chp} are integrals of motion (and some of them are
Casimirs in the even $n$ cases), which Poisson commute according to our general reasoning from sect.~\ref{ss:isrmatr}. The genus of the curve \rf{chp} is always
\be
\label{genus}
g=2\left(\left[{N+1\over 2}\right]-1\right)=\left\{\begin{array}{c}
2\left[{N\over 2}\right],\ \ \ N\ {\rm odd} \\
 2\left(\left[{N\over 2}\right]-1\right) ,\ \ \ N\ {\rm even}
                                                   \end{array}\right.
\ee
which can be easily seen from the corresponding Newton polygon at fig.~\ref{fi:newtonOT}.

\noindent
{\bf Examples}. For $N=3$
\be
\label{Cn3}
C_2(\lambda) = \lambda^3+\lambda^2\mathcal{H}_2, \ \ \ \
C_1(\lambda) = -1 + \lambda \mathcal{H}_1
\ee
with the Poisson commuting Hamiltonians ($X_0X_1X_2=1$, $Y_0Y_1Y_2=1$)
\be
\label{Rhamschw}
\mathcal{H}_1 = X_1+X_2+X_1X_2Y_1+ {Y_2\over X_1}+{1\over X_1X_2}+{1\over X_2Y_1Y_2}
\\
\mathcal{H}_2 = Y_1+Y_2+X_2Y_1Y_2+ {X_1\over Y_2}+{1\over Y_1Y_2}+{1\over X_1X_2Y_1}
\ee
w.r.t.
\be
\label{pbxyred}
\{ \log X_1,\log X_2\} = 1 = \{ \log Y_2,\log Y_1\}
\ee
In symmetric form \rf{chp} can be rewritten as
\be
\label{specuRT}
w+{1\over w} = z^{-3/2}P_3(z),\ \ \ \ w=(-\mu)^{1/2}\lambda^{3/2},\ z=-\mu\lambda
\\
P_3(z) = z^3-\mathcal{H}_2z^2+\mathcal{H}_1z -1
\ee
and acquires the form of spectral curve \rf{specu} for the relativistic $\widehat{SL(3)}$ Toda chain\footnote{To avoid misunderstanding, let us point out, that for $N=3$ the spectral variables
in equations \rf{chapo} and \rf{chp} are related by $\lambda\leftrightarrow\lambda$, but $\mu\leftrightarrow\mu\cdot\lambda$. This transformation maps the Newton polygons from
fig.~\ref{fi:newtod} and
fig.~\ref{fi:newtonOT} to each other.}.

Re-writing \rf{Rhamschw} in the cluster co-ordinates $y_{i+3}=y_i$, $x_{i+3}=x_i$, using
\be
\label{xyuv}
X_1=y_1\prod_{k=1,2} \left(y_kx_k\right)^{-C^{-1}_{1k}}, \ \ \ \
X_2=y_1x_1y_2\prod_{k=1,2} \left(y_kx_k\right)^{-C^{-1}_{1k}}
\\
Y_1=x_1x_2\prod_{k=1,2} \left(y_kx_k\right)^{-C^{-1}_{2k}},\ \ \ \
Y_2=y_1y_2x_2\prod_{k=1,2} \left(y_kx_k\right)^{-C^{-1}_{2k}}
\ee
one comes back to the formulas \rf{Rhamsl3p}. For $N=4$ one gets from \rf{chp}
\be
C_2(\lambda) = \lambda^4 + \lambda^3\mathcal{H}_2+\lambda^2K_2({\bf Y})
\\
C_1(\lambda) = 1 - \lambda \mathcal{H}_1 +\lambda^2K_2({\bf X})
\ee
with $K_2({\bf X})=X_1X_3+{1\over X_1X_3}$ and the same for $K_2({\bf Y})$, while
\be
\mathcal{H}_1=X_1+X_2+X_3+X_2X_1Y_1+X_3X_2Y_2+{Y_3\over X_1X_2}+{1\over Y_1Y_2Y_3X_2X_3}+{1\over X_1X_2X_3}=
\\
=X_1+{1\over X_1}+X_2+{1\over X_2}+X_1X_2Y_1+{1\over X_1X_2Y_1}+{X_2Y_2\over X_1}+{X_1\over Y_2X_2}
\\
\mathcal{H}_2=Y_1+Y_2+Y_3+Y_1Y_2X_2+Y_2Y_3X_3+{X_1\over Y_2Y_3}+{1\over X_1X_2X_3Y_1Y_2}+{1\over Y_1Y_2Y_3} =
\\
= Y_1+{1\over Y_1}+Y_2+{1\over Y_2}+Y_1Y_2X_2+{1\over Y_1Y_2X_2}
+{Y_2\over X_1Y_1}+{X_1Y_1\over Y_2}
\ee
where the r.h.s.'s are obtained by fixing $K_2({\bf X})=K_2({\bf Y})=2$. These Hamiltonians do
Poisson commute w.r.t. the same bracket \rf{pbxyred}, as in the $N=3$ case. The spectral curve equation
\rf{detT} turns to be
\be
\label{specun4}
\mu^3\lambda^4 + \mu^2(\lambda^4 + \lambda^3\mathcal{H}_2+2\lambda^2)+\mu(1 - \lambda \mathcal{H}_1 +2\lambda^2)+1 =0
\ee
For $N\geq 5$ these formulas give rise to integrable system, related with the pentagram map \cite{OT}.
Basically, \rf{xiopp3} turns to be the spectral parameter dependent Lax matrix for this system, proposed
in \cite{FockP} by relating it to the co-extended loop group (another Lax representation was found in \cite{Sol,GeShaLast}).

\setcounter{equation}0
\section{Discussion
\label{ss:disc}}

We have demonstrated that relativistic Toda systems \cite{Ruj} arise naturally
on Poisson submanifolds in Lie groups. Extension of the construction from simple
Lie groups to the co-extended loop groups not only gives rise to
their periodic versions, but allows to consider much
wider class of integrable models \cite{FM11,FM12}. We have discussed in previous section
just few particular examples of them, whose phase spaces can be identified with
the Poisson manifolds, obtained by regular gluing of two-dimensional
square lattice.

The proposed construction of the Poisson
submanifolds in loop groups gives rise their spectral curve equations and
explicit formulas for the coefficients - the integrals of motion (corresponding usually to the
internal points of their Newton polygons) in cluster variables. These formulas
have similar structure with partition functions of the dimer models,
and this is not a coincidence. There is indeed a direct relation between the class of
integrable systems, coming from the dimer models on bipartite graphs on a torus,
considered in \cite{GK}, and those constructed from affine Lie groups: for any convex Newton polygon
one can construct a wiring diagram on torus, and after cutting the torus - an element from the co-extended
double Weyl group \cite{FM12}. On one hand this gives a Poisson submanifold in loop group, and an integrable system by the Lax map described above, on another hand each wiring diagram gives a bipartite graph
on torus, so that the face dimer partition function produce the same integrals of motion as a coefficients
of the spectral curve equation.

We have also left beyond the scope of this paper the quantum versions of our integrable models.
It has been notices already in \cite{FM} that
quantization of the Toda chain can be easily done within the proposed approach by passing from classical to
quantum groups, since the problem is identical
to the problem of deformation of the algebra of functions $Fun(G)$ on group $G$ \cite{QG}.
Notice only, that in cluster language the quantisation looks especially
simple, since the Poisson brackets \rf{pbclust} are always logarithmically constant.

Finally let us point out, that representatives of the family of integrable systems we
consider in the paper found to have many applications in mathematical physics, in
particular being responsible for the Seiberg-Witten exact
solutions to the theories with explicitly present compact extra dimensions
\cite{SWcompact}. We expect a nontrivial relation between the cluster formulation of these
integrable models and recently discussed cluster mutations, arising as generating
transformations for the BPS-spectra, and being related to the wall-crossing phenomenon (see e.g. \cite{wallcross}).
We are going to return to these issues elsewhere.

\section*{Acknowledgements}

I am grateful to A.~Braverman, B.~Dubrovin, N.~Early, S.~Kharchev, S.~Khoroshkin, I.~Krichever, O.~Kruglinskaya,
A.~Morozov, P.~Pyatov, P.~Saponov, M.~Shapiro and A.~Zabrodin for useful discussions and
to V.~Fock for the fruitful collaboration during many years of thinking on the subject of these notes.
I would like to thank the University of Aarhus, the Max Planck and Haussdorf Institutes for Mathematics in Bonn and University of Strasbourg, where essential parts of this work have been done. The work was partially supported by the RFBR grant 11-01-00962, by the joint RFBR projects 10-02-92108 and 11-01-92612,
by the program of support of scientific schools NSh-3349.2012.2 and by the State contract 14.740.11.0081.

\appendix

\section*{Appendix}
\setcounter{equation}0
\section{Roots, weights, Cartan matrices
\label{ap:promunu}}

For the $\mathfrak{g}=sl_N$ Lie algebras the Cartan matrix is
\be
\label{CslN}
C_{ij}\ \stackreb{sl_N}{=}\ (\alpha_i\cdot\alpha_j)= 2\delta_{ij} - \delta_{i+1,j}-\delta_{i,j+1}
\ee
for the positive simple roots $\alpha_i\in\Pi$, $i,j=1,\ldots,\rank\ G=N-1$.
The dual vectors
\be
\label{amu}
(\mu_i\cdot\alpha_j)=\delta_{ij}
\ee
are the highest weights of the fundamental representations. Clearly,
\be
\label{aCmu}
\alpha_i = \sum_j C_{ij}\mu_j,\ \ \ \
\mu_i = \sum_j C^{-1}_{ij}\alpha_j
\ee
It is convenient to use the weight-vectors of the first fundamental representation
\be
\label{numu}
\nu_i = \mu_i-\mu_{i-1},\  \ \ \ i=1,\ldots,N
\ee
where it is implied that $\mu_0=\mu_N=0$ (i.e. $\nu_1=\mu_1$ and $\nu_N=-\mu_{N-1}$), with
the scalar products
\be
\label{nunu}
(\nu_i\cdot\nu_j) = \delta_{ij}-{1\over N},\ \ \ \ i,j=1,\ldots,N
\ee
Vectors \rf{numu} are easily presented as
\be
\label{nue}
\nu_i = {\bf e}_i - {1\over N}\sum_{j=1}^N {\bf e}_j,\ \ \ i=1,\ldots,N
\ee
projection of the Cartesian basis vectors in $N$-dimensional vector space to a hyperplane, orthogonal
to the vector ${1\over N}\sum_{j=1}^N {\bf e}_j$. Comparing \rf{aCmu} and \rf{numu}, and using \rf{nue} one finds that
\be
\label{anu}
\alpha_i = \nu_i-\nu_{i+1} = {\bf e}_i-{\bf e}_{i+1},\ \ \ \ i=1,\ldots,N-1
\ee
Recall that given a Cartan matrix $C_{ij}$, the Lie algebra $\mathfrak{g}$ is generated by $\{h_i|i \in \Pi\}$ and $\{e_i|i \in \Pi\cup{\bar\Pi}\}$ subject to the relations\footnote{To simplify them we extend $h$ and $C$ to negative values of indices, assuming that $h_i = h_{-i}$ and that $C_{-i,-j}=C_{ij}$ and $C_{ij}=0$ if $i$ and $j$ have different signs. We shall also use the notation $e_{-i}=e_{\bar i}$ for $i>0$.}
\be\label{rela}
[h_i,h_j]=0,\ \ \ [h_i,e_{j}] = \sign(j)C_{ij} e_j,\ \ \ [e_i,e_{-i}] =\sign(i) h_i,
\\
(\Ad\ e_i)^{1-C_{ij}}e_j = 0 \mbox{ for } i+j\neq 0
  \ee
One can also replace the set $\{ h_i\}$ by the dual set $\{h^i\}$ by $h_i=\sum_{j\in\Pi}C_{ij}h^j$, so
that
\be\label{relx}
[h^i,h^j]=0,\ \ \ [h^i,e_{j}] = \sign(j)\delta_i^j e_j,\ \ \ [e_i,e_{-i}] =\sign(i) C_{ij}h^j,
\\
(\Ad\ e_i)^{1-C_{ij}}e_j = 0 \mbox{ for } i+j\neq 0
  \ee
For any $i \in \Pi\cup{\bar\Pi}$ one can introduce the group element $E_i = \exp(e_i)$ and a one-parameter subgroup $H_i(z)=\exp(\log z\cdot h^{i})$.

In the case of affine Lie algebra $\mathfrak{g}=\widehat{sl}_N$ the set of simple roots is extended
$\alpha_i\in\Pi = \mathbb{Z}_N$ in cyclically symmetric way
and the Cartan matrix
\be
\label{CslNh}
\hat{C}_{ij}\ \stackreb{\widehat{sl}_N}{=}\ 2\delta_{ij} - \delta_{i+1,j}-\delta_{i,j+1},\ \ \ i,j \in \mathbb{Z}_N
\ee
is degenerate.

\setcounter{equation}0
\section{r-matrices and Yang-Baxter equations
\label{ap:rmatr}}

The classical limit of the Yang-Baxter equation
\be
\label{YBE}
\hat{R}_{12}\hat{R}_{13}\hat{R}_{23}=\hat{R}_{23}\hat{R}_{13}\hat{R}_{12}
\ee
for the $\hat{R}_{ij}=\exp\left(\hbar\ \hat{r}_{ij}\right)$ at $\hbar\to 0$ has the form
\be
\label{CYBE}
\left[\hat{r}_{12},\hat{r}_{13}\right] + \left[\hat{r}_{12},\hat{r}_{23}\right] + \left[\hat{r}_{13},\hat{r}_{23}\right]=0
\ee
It can be solved, for example, by
\be
\label{rp}
\hat{r} = r_+ = \sum_{\alpha\in\Delta_+}\left( e_\alpha\otimes e_{\bar\alpha} + {\bf h}\otimes {\bf h}\right)\ \stackreb{sl_2}{=}\  e\otimes {\bar e} + \quatr h\otimes h
\ee
where (in the fundamental representation of $sl_2$)
\be
\label{efh}
e = \left(\begin{array}{cc}
            0 & 1 \\
            0 & 0
          \end{array}\right),\ \ \
{\bar e} = \left(\begin{array}{cc}
            0 & 0 \\
            1 & 0
          \end{array}\right),\ \ \
h = \left(\begin{array}{cc}
            1 & 0 \\
            0 & -1
          \end{array}\right)
\ee
as well as by
\be
\label{rm}
\hat{r} = r_- = \sum_{\alpha\in\Delta_+}\left( e_{\bar\alpha}\otimes e_\alpha + {\bf h}\otimes {\bf h}\right)\ \stackreb{sl_2}{=}\  {\bar e}\otimes e + \quatr h\otimes h
\ee
obtained from \rf{rp} by involution $e_\alpha\leftrightarrow e_{\bar\alpha}$,  $\alpha\in\Delta_+$ and ${\bf h}\leftrightarrow -{\bf h}$,
preserving the commutation relations. The anti-symmetric $r$-matrix
 \be
\label{rmat2}
r=e\otimes {\bar e}-{\bar e}\otimes e
=\left(\begin{array}{cccc}
 0  & 0 & 0 & 0 \\
 0  & 0 & 1 & 0 \\
 0  & -1 & 0 & 0\\
 0 & 0 & 0 & 0
\end{array}\right) =r_+-r_-
\ee
satisfies the modified classical Yang-Baxter equation: for example, explicit calculation for \rf{rmat2} gives
\be
\label{mYBE}
\left[r_{12},r_{13}\right] + \left[r_{12},r_{23}\right] + \left[r_{23},r_{13}\right] =
h\wedge e\wedge {\bar e}
\ee
which is enough to guarantee the Jacobi identity for the Poisson bracket \rf{rbra}.

In the case of loop algebras the corresponding $\hat{r}$-matrices acquire the spectral parameter
dependence, corresponding to the evaluation representations of the corresponding current algebras (see
e.g. \cite{arut}).
Direct application of the anti-symmetric formula \rf{ref} for the $\mathfrak{g}=\widehat{sl}_2$ gives rise to
\be
\label{rhsl2}
\sum_{\alpha\in \Delta_+} e_\alpha\wedge e_{\bar\alpha} =
 \sum_{n\ge 0}e_n\wedge {\bar e}_{-n} + \sum_{n\ge 1}\left({\bar e}_n\wedge e_{-n} +
 \half h_n\wedge h_{-n}\right)=
\\
= - {\lambda +\lambda '\over \lambda -\lambda '}\left(e\otimes {\bar e} + {\bar e}\otimes e + \half h\otimes h\right)
+ e\otimes {\bar e} - {\bar e}\otimes e =
\\
= - {\lambda +\lambda '\over \lambda -\lambda '}\
\left(\begin{array}{cccc}
        \half & 0 & 0 & 0 \\
        0 & -\half & 1 & 0 \\
        0 & 1 & -\half & 0 \\
        0 & 0 & 0 & \half
      \end{array}\right) + \left(\begin{array}{cccc}
 0  & 0 & 0 & 0 \\
 0  & 0 & 1 & 0 \\
 0  & -1 & 0 & 0\\
 0 & 0 & 0 & 0
\end{array}\right) =
\\
= \left(\begin{array}{cccc}
 0  & 0 & 0 & 0 \\
 0  & {\lambda +\lambda '\over \lambda -\lambda '} & -{2\lambda '\over \lambda -\lambda '} & 0 \\
 0  & -{2\lambda \over \lambda -\lambda '} & {\lambda +\lambda '\over \lambda -\lambda '} & 0\\
 0 & 0 & 0 & 0
\end{array}\right) - {1\over 2}{\lambda +\lambda '\over \lambda -\lambda '}\ {\bf 1}\otimes {\bf 1}
\ee
Expression in the r.h.s., up to the part proportional to the unity operator, which
does not give contribution to any commutators, acquires the form of
\be
\label{rtrig}
r_{\rm trig}(\lambda ,\lambda'  ) = \left(\begin{array}{cccc}
 0  & 0 & 0 & 0 \\
 0  & {\lambda +\lambda'  \over \lambda -\lambda'  } & -{2\lambda'  \over \lambda -\lambda'  } & 0 \\
 0  & -{2\lambda \over \lambda -\lambda'  } & {\lambda +\lambda'  \over \lambda -\lambda'  } & 0\\
 0 & 0 & 0 & 0
\end{array}\right)
\ee
arising in the Poisson bracket relations \rf{PbS} in the context of $2\times 2$ formalism. In the
rational limit, with $\lambda =i\exp(\zeta/2)$ and small $\zeta$, one gets from \rf{rtrig}
\be
\label{rtrlim}
r_{\rm trig}\left(ie^{\zeta/2},ie^{\zeta'/2}\right)\ \stackreb{\zeta,\zeta'\to 0}{=}\
{2\over\zeta-\zeta'}\left(\begin{array}{cccc}
 0  & 0 & 0 & 0 \\
 0  & 1 & -1 & 0 \\
 0  & -1 & 1 & 0\\
 0 & 0 & 0 & 0
\end{array}\right) + \ldots
\ee
Adding to the r.h.s. the term $-{2\over\zeta-\zeta'}{\bf 1}\otimes{\bf 1}$, one gets in this limit $r_{\rm trig}\left(ie^{\zeta/2},ie^{\zeta'/2}\right)\ \stackreb{\zeta,\zeta'\to 0}{=}\ -2r_{\rm rat}(\zeta,\zeta')$,
where
\be
\label{rrat}
r_{\rm rat}(\zeta,\zeta') = {1\over\zeta-\zeta'}\left(\begin{array}{cccc}
 1  & 0 & 0 & 0 \\
 0  & 0 & 1 & 0 \\
 0  & 1 & 0 & 0\\
 0 & 0 & 0 & 1
\end{array}\right) = {1\over\zeta-\zeta'}\left(\half\left({\bf 1}\otimes{\bf 1} +  h\otimes h\right) + e\otimes {\bar e}+{\bar e}\otimes e\right)
\ee
is the rational $r$-matrix, proportional to the permutation operator in the tensor product $\mathbb{C}^2\otimes\mathbb{C}^2$.

\setcounter{equation}0
\section{Poisson structure on group manifold
\label{ap:poisson}}

\subsection{Graphs and Poisson brackets}

Consider a graph (or a quiver) $\Gamma$ - a set of $|\Gamma|$ vertices, connected by arbitrary number of the oriented edges. Assign to each vertex $I\in\Gamma$ a (complex or real) variable $z_I$, $I=1,\ldots,|\Gamma|$, to be thought of as co-ordinates on a chart in some manifold, mapped into $\left(\mathbb{C}^\times\right)^{|\Gamma|}$. On such open chart one can define the Poisson bracket \rf{pbclust} (see, e.g. \cite{Drinfeld})
\be
\label{clbra}
\{ z_I,z_J\} = \varepsilon_{IJ}z_Iz_J,\ \ \ \ I,J=1,\ldots,|\Gamma|
\ee
where $\varepsilon_{IJ}$ stays for the exchange matrix
\be
\label{ince}
\varepsilon_{IJ} = \# {\rm arrows}\ (I\rightarrow J) - \# {\rm arrows}\ (J\rightarrow I)
\ee
Obviously, $\varepsilon_{IJ}=-\varepsilon_{JI}$, and the Jacobi identity is satisfied for \rf{clbra} automatically. Moreover, exchange matrix \rf{ince} can be even non-integer (e.g. half-integer) valued.

The so defined Poisson manifolds (see \cite{Drinfeld} and references therein for details) allow several operations, preserving the Poisson structure:

\begin{itemize}
  \item For any subset $\Gamma'\subset\Gamma$ one can put
\be
\label{forget}
\varepsilon_{IJ'}=0,\ \ \ \forall\ I\in\Gamma,\ \ \forall\ J'\in\Gamma'
\ee
which corresponds just to ``forgetting'' all vertices of $\Gamma$ with the variables $\{ z_{J'}\}$, $J'\in\Gamma'$.
  \item Gluing: for two graphs $\Gamma_1$ and $\Gamma_2$ {\em identify} their subsets
  $\Gamma'_1=\Gamma'_2=\Gamma'$, getting a graph
\be
\label{glue}
  \Gamma=\Gamma_1\cup\Gamma_2 = \Gamma_1\diagdown\Gamma'_1\ \cup \Gamma'\ \cup \Gamma_2\diagdown\Gamma'_2
\ee
with the variables $z_{I_1}=z^{(1)}_{I_1}$ in all vertices $I_1\in \Gamma_1\diagdown\Gamma'_1$,
$z_{I_2}=z^{(2)}_{I_2}$ for $I_2\in \Gamma_2\diagdown\Gamma'_2$, and $z_{I'}=z^{(1)}_{I'}z^{(2)}_{I'}$ for the coinciding $I'\in \Gamma'$. One gets then for the exchange matrix of \rf{glue}
\be
\label{incglue}
\varepsilon_{I_kJ}=\varepsilon^{(k)}_{I_kJ},\ \ \ \ J\in\Gamma,\ \ I_k\in\ \Gamma_k\diagdown\Gamma'_k,\ \ k=1,2
\\
\varepsilon_{I'J'}=\varepsilon^{(1)}_{I'J'}+\varepsilon^{(2)}_{I'J'},\ \ \ \ I',J'\in\ \Gamma'
\ee
  \item The Poisson structure \rf{clbra} (for {\em integer-valued} $\varepsilon_{ij}$) is preserved by {\em mutations} of the graph and corresponding transformation of the $z$-variables
\be
\label{mutz}
\mu_J:\ \  z_J\rightarrow {1\over z_J},\ \ \ \ \  z_I\rightarrow z_I\left(1+z_J^{{\rm sgn}(\varepsilon_{IJ})}\right)^{\varepsilon_{IJ}}, I\neq J
\ee
which allow to extend the bracket from an open chart to some globally defined cluster variety \cite{cv}. \end{itemize}
For certain graphs the Poisson structure \rf{clbra} becomes equivalent
to the $r$-matrix Poisson structure \rf{rbra} on the Lie groups, since it turns to be possible
to define the group-theoretical
multiplication on certain graphs \cite{Drinfeld}. We shall illustrate it here on few particular examples.

\subsection{Graphs and groups}

\begin{figure}[tb]
\epsfysize=3cm
\centerline{\epsfbox{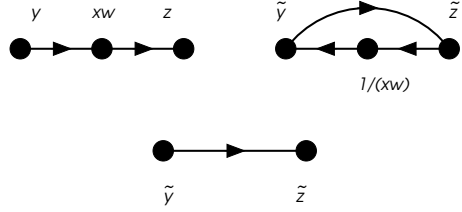}}
\caption{\sl Simplest graph and Lie group. Gluing, mutation at the middle point and forgetting this vertex after mutation gives rise to the group multiplication law of the upper-triangular subgroup of $SL(2)$.}
\label{fi:triang}
\end{figure}

\paragraph{1.} The simplest graph $y\rightarrow x$ corresponds to the subgroup of upper-triangular matrices in $SL(2)$ or $PGL(2)$ (sometimes it is easier just to forget about the determinant). More strictly, consider it as a short-hand notation for
\be
\label{1arrow}
y\stackreb{E}{\longrightarrow} x = YEX = \left(\begin{array}{cc}
                                             y & 0 \\
                                             0 & 1
                                           \end{array}\right)\left(\begin{array}{cc}
                                             1 & 1 \\
                                             0 & 1
                                           \end{array}\right)\left(\begin{array}{cc}
                                             x & 0 \\
                                             0 & 1
                                           \end{array}\right) = \left(\begin{array}{cc}
                                             yx & y \\
                                             0 & 1
                                           \end{array}\right)
\ee
with the matrix in r.h.s. considered as an element of $PGL(2)$. Consider a multiplication $(y\rightarrow x)\cdot (w\rightarrow z) = y\rightarrow xw\rightarrow z$, according to the gluing rule. After mutating at the intermediate point, the variables according to \rf{mutz} transform as
\be
\label{muyw}
xw \rightarrow {1\over xw},\ \ \
y \rightarrow y(1+xw)={\tilde y},\ \ \
z\rightarrow z\left(1+{1\over xw}\right)^{-1}={\tilde z}
\ee
which corresponds to the second graph from fig.~\ref{fi:triang},
with an extra arrow between ${\tilde y}$ and ${\tilde z}$ reflecting, that
\be
\{ {\tilde y},{\tilde z}\} = \left\{y(1+xw),z\left(1+{1\over xw}\right)^{-1}\right\} =
\\
=(1+xw)z\left\{y,\left(1+{1\over xw}\right)^{-1}\right\} +
y\left(1+{1\over xw}\right)^{-1}\left\{(1+xw),z\right\} =
\\
= yxwz = {\tilde y}{\tilde z}
\ee
Forgetting the intermediate vertex, as at fig.~\ref{fi:triang}, one gets the original graph
${\tilde y}\rightarrow {\tilde z}$ but with the new variables in the vertices. The rule to get them exactly corresponds to
\be
\label{1arprod}
\left(y\stackreb{E}{\longrightarrow} x\right)\cdot\left( w\stackreb{E}{\longrightarrow} z\right) =  \left(\begin{array}{cc}
yx & y \\
0 & 1
\end{array}\right)\left(\begin{array}{cc}
wz & w \\
 0 & 1
 \end{array}\right) =
\left(\begin{array}{cc}
yxwz & y(1+xw) \\
 0 & 1
  \end{array}\right) =
\\
= \left(\begin{array}{cc}
{\tilde y}{\tilde z} & {\tilde y} \\
 0 & 1
  \end{array}\right) = {\tilde Y}E{\tilde Z} = {\tilde y}\ \stackreb{E}{\longrightarrow}\ {\tilde z}
\ee
multiplication of the upper-triangular matrices. Normalising \rf{1arrow} to be an element of $SL(2)$
\be
\label{1arsl2}
y\stackreb{E}{\longrightarrow} x = YEX = \left(\begin{array}{cc}
                                             y^{1/2} & 0 \\
                                             0 & y^{-1/2}
                                           \end{array}\right)\left(\begin{array}{cc}
                                             1 & 1 \\
                                             0 & 1
                                           \end{array}\right)\left(\begin{array}{cc}
                                             x^{1/2} & 0 \\
                                             0 & x^{-1/2}
                                           \end{array}\right) =
\\
 = \left(\begin{array}{cc}
 \sqrt{yx} & \sqrt{y\over x} \\
   0 & {1\over\sqrt{yx}}
    \end{array}\right) \in SL(2)
\ee
one finds that the Poisson bracket $\{ y,x\} = yx$ just coincides with the restriction of $SL(2)$ r-matrix Poisson bracket \rf{rbra} to the subgroup, generated by
the only positive root $E=\exp(e)$.

\begin{figure}[hc]
\epsfysize=3cm
\centerline{\epsfbox{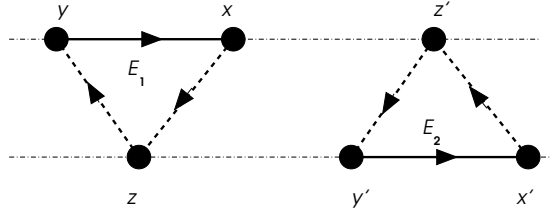}}
\caption{\sl Exchange graphs, corresponding to the subgroups of $SL(3)$ generated by positive simple roots $1$ (left) and $2$ (right). The variables at the vertices on upper level parameterise the Cartan subgroup $H_1$, and from the lower level - generated by $H_2$.}
\label{fi:treug}
\end{figure}

\paragraph{2.} Consider now \rf{1arrow} as an upper-triangular subgroup of $SL(3)$, generated by any of the simple roots $E_i=\exp(e_i)$, and the  Cartan elements $H_i(z)=\exp(\log z\cdot h^i)$, $i=1,2$.
Using \rf{ref} one can easily compute the corresponding Poisson brackets, e.g. for the matrix elements of
\be
\label{3bra1}
H_1(y)E_1H_1(x)H_2(z)=\left(
\begin{array}{ccc}
  y^{2/3} & 0 & 0 \\
  0 & y^{-1/3} & 0 \\
  0 & 0 & y^{-1/3}
\end{array}\right)\cdot
\\
\cdot\left(
\begin{array}{ccc}
  1 & 1 & 0 \\
  0 & 1 & 0 \\
  0 & 0 & 1
\end{array}\right)\left(
\begin{array}{ccc}
  x^{2/3} & 0 & 0 \\
  0 & x^{-1/3} & 0 \\
  0 & 0 & x^{-1/3}
\end{array}\right)
\left(
\begin{array}{ccc}
  z^{1/3} & 0 & 0 \\
 0  & z^{1/3} & 0 \\
  0 & 0 & z^{-2/3}
\end{array}\right) =
\\
= \left(
\begin{array}{ccc}
  y^{2/3}x^{2/3}z^{1/3} & y^{2/3}z^{1/3}x^{-1/3} & 0 \\
  0 & z^{1/3}y^{-1/3}x^{-1/3} & 0 \\
  0 & 0 & y^{-1/3}x^{-1/3}z^{-2/3}
\end{array}\right)
\ee
one gets the following Poisson relations
\be
\label{pbtreug}
\{ y,x\} = yx,\ \ \ \{x,z\} = \half xz,\ \ \ \{ z,y\} = \half zy
\ee
which can be encoded in the left graph from fig.~\ref{fi:treug}, with
the {\em half}-arrows, connecting $z$ with $y$ and $x$.

Similarly, for the subgroup, generated by the second root
\be
H_2(y')E_2H_2(x')H_1(z') = H_1(z')H_2(y')E_2H_2(x')
\ee
one gets the right graph from fig.~\ref{fi:treug},
producing the same Poisson relations \rf{pbtreug} for the prime-variables.
Almost the same triangles correspond to the subgroups, generated by the negative roots $E_{\bar 1}$ and $E_{\bar 2}$ - one has only to change the orientation of all arrows on these pictures.

\begin{figure}[tb]
\epsfysize=4cm
\centerline{\epsfbox{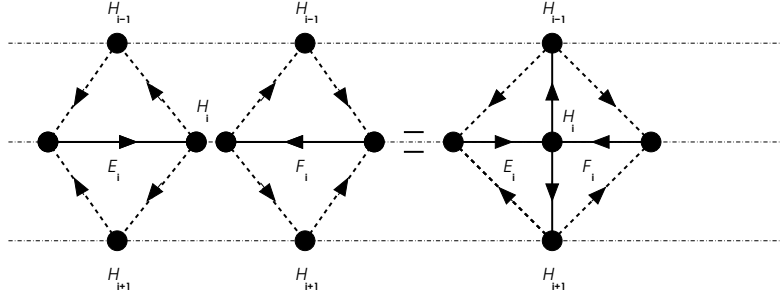}}
\caption{\sl Graphs, depicting the subgroups of $SL(N)$ generated by the simple positive root $E_i$ and negative root $E_{\bar i}$, $1<i<N-1$. The variables at the vertices on each level correspond to the Cartan generators $H_i$, and $H_{i\pm 1}$ correspondingly. Gluing two such graphs one gets an element of the two-dimensional square lattice.}
\label{fi:rhomb}
\end{figure}

The same arguments show, that for simple roots of $SL(N)$ the corresponding subgroups
are generated by {\em rhombi} instead of triangles, see fig.~\ref{fi:rhomb}.
The half-arrows on these rhombi are of the same nature as at fig.~\ref{fi:treug}, coming
from the half-integer coefficients in \rf{rbra}.
Gluing two such rhombi at $i$-th level one gets an element of the square lattice, see fig.~\ref{fi:rhomb}.
The symplectic leaves in $SL(N)$ can be constructed by further gluing of the elements from fig.~\ref{fi:rhomb}, see fig.~\ref{fi:lattice}. In such way one gets some particular gluing
of the two-dimensional square lattice, among them is the Toda symplectic leave in $SL(N)$ of
dimension $2\cdot\rank\ SL(N) = 2(N-1)$.

\setcounter{equation}0
\section{Toda theory from Lie algebra
\label{ap:alg}}

In the case of Lie algebra instead of \rf{rbra} one has the linear Poisson bracket
\be\label{rbral}
\left\{ {\cal L} \stackreb{,}{\otimes}{\cal L} \right\} = -\half \left[r,({\cal L}\otimes 1+1\otimes {\cal L})\right]
\ee
with the same constant $r$-matrix \rf{ref}.
In the case of $GL(2)$ or $SL(2)$ it has only two non vanishing entries \rf{rmat2},
giving the following Poisson bracket relations for ${\cal L}=\left(\begin{array}{cc}
                                                            a & b \\
                                                            c & d
                                                          \end{array}\right)\in gl_2$
\be
\{ a,a \} = 0,\ \ \ \{ a,b \}=-\half b, \ \ \ \{ a,c \}=-\half c,\ \ \ \{ a, d \}=0
\\
\{ b,c \}=0,\ \ \ \{ b,d \}=-\half b,\ \ \ \{ c,d\}=-\half c
\ee
which reflect the structure of the dual Lie algebra ($\Tr{\cal L}=a+d$ is now the Casimir function - the total momentum).
The Darboux variables are introduced via
\be
{\cal L}=\left(\begin{array}{cc}
a & b \\
c & d
\end{array}\right) =
\left(\begin{array}{cc}
p & e^{q/2} \\
e^{q/2} & -p
\end{array}\right)\in sl_2,\ \ \ \ \{ q, p\}=1
\ee
and the canonical Hamiltonian is
given by $H=H_2=\half\Tr\ {\cal L}^2 = p^2+e^{q}$.

Generally, the Lie-algebraic valued matrix of the form
\be
\label{gtl}
{\cal L} =  (p\cdot h) + \sum_{i\in\Pi}
\left(e_i+e_{\bar i}\right)\exp(\alpha_i\cdot q)
\ee
satisfies \rf{rbral} and produces the Hamiltonians of the Toda system in a standard way -
by computing invariant functions of \rf{gtl}. If the sum in \rf{gtl} is taken over the set $\Pi$ of positive simple roots for the finite-dimensional Lie algebra one gets the open Toda chain, while for the affine Lie algebra with $e_i$'s and $e_{\bar i}$'s taken in the evaluation
representation one gets the spectral parameter dependent Lax operator for the periodic
Toda chain.

In the context of $2\times 2$ formalism, the Lax matrix \rf{L22} turns in the algebraic limit (roughly: linear in momenta, exponentiated co-ordinates
and $\zeta$, if $\lambda  =i\exp(\zeta/2)$), into
\be
\label{L22nr}
{1\over i}\left(\begin{array}{cc}
                      0 & {e^{-P_j/2-q_j}\over \sqrt{\lambda  }} \\
                      \sqrt{\lambda  }e^{P_j/2+q_j}\ \ & \ \sqrt{\lambda  }e^{-P_j/2}+{e^{P_j/2}\over \sqrt{\lambda  }}
                    \end{array}\right)\rightarrow
                    L_j(\zeta) = \left(\begin{array}{cc}
                      0 & -\exp\left(-q_j/2\right) \\
                      \exp\left(q_j/2\right)\ \ & \ \zeta-p_j
                    \end{array}\right)
                    \\
j=1,\ldots,N
\ee
which satisfies the Poisson bracket  \cite{FT}
\be
\label{PbSnr}
\left\{ L_i(\zeta) \stackreb{,}{\otimes}L_j(\zeta') \right\}= \delta_{ij}\left[r_{\rm rat}\left(\zeta,\zeta'\right), L_i(\zeta) \stackreb{,}{\otimes}L_j(\zeta')\right]
\ee
with the rational r-matrix \rf{rrat}.

\end{document}